\documentclass{svmult}

\usepackage{epsfig}
\usepackage{bm}                 
\usepackage{amsmath}
\usepackage{amssymb}
\usepackage{graphicx}
\usepackage{psfrag}
\usepackage{footmisc}

\usepackage[matrix,frame,arrow]{xy}
\usepackage{amsmath}

\newcommand{\ket}[1]{\left\vert{#1}\right\rangle}
\newcommand{\qw}[1][-1]{\ar @{-} [0,#1]}
\newcommand{\qwx}[1][-1]{\ar @{-} [#1,0]}

\newcommand{\gate}[1]{*{\xy *+<.6em>{#1};p\save+LU;+RU **\dir{-}\restore\save+RU;+RD **\dir{-}\restore\save+RD;+LD **\dir{-}\restore\POS+LD;+LU **\dir{-}\endxy} \qw}
\newcommand{\meter}{\gate{\xy *!<0em,1.1em>h\cir<1.1em>{ur_dr},!U-<0em,.4em>;p+<.5em,.9em> **h\dir{-} \POS <-.6em,.4em> *{},<.6em,-.4em> *{} \endxy}}

\newcommand{\control}{*-=-{\bullet}}

\newcommand{\ctrl}[1]{\control \qwx[#1] \qw}

\newcommand{\targ}{*{\xy{<0em,0em>*{} \ar @{ - } +<.4em,0em> \ar @{ - } -<.4em,0em> \ar @{ - } +<0em,.4em> \ar @{ - } -<0em,.4em>},*+<.8em>\frm{o}\endxy} \qw}
\newcommand{\rstick}[1]{*!L!<-.5em,0em>=<0em>{#1}}
\newcommand{\lstick}[1]{*!R!<.5em,0em>=<0em>{#1}}

\newcommand{\Qcircuit}{\xymatrix @*=<0em>}

\def\vac{\circledcirc}
\def\unity{\mathbb{I}}
\def\down{\text{\mbox{$\uparrow$}}}
\def\up{\text{\mbox{$\downarrow$}}}

\def\cline#1#2{\noindent{\bf \ref{lec:#2}~ #1}\hfill\pageref{lec:#2} \smallskip}
\def\cdes#1{\parbox{11cm}{#1}\medskip}

\usepackage[colorlinks=false]{hyperref}

\begin{document}

\title*{Five Lectures on \\ Optical Quantum Computing}
\author{Pieter Kok}
\institute{Quantum \& Nano-Technology Group, Department of Materials, Oxford University, Parks Road, Oxford OX1 3PH, United Kingdom.}
\titlerunning{Optical Quantum Computing}
\maketitle


\cline{Light and quantum information}{light}\\
\cdes{Photons as qubits, phase shifters, beam splitters, polarization rotations, polarizing beam splitters, interferometers.}\\
\cline{Two-qubit gates and the KLM scheme}{klm}\\
\cdes{Two-photon entanglement, the KLM approach, Clifford operations, two-photon interference, Hong-Ou-Mandel effect, fusion gates.}\\
\cline{Cluster states}{cluster}\\
\cdes{From circuits to clusters, single-qubit gates, two-qubit gates, universal cluster states, making clusters with fusion gates.}\\
\cline{Quantum computing with matter qubits and photons}{dh}\\
\cdes{Quantum memories, double-heralding entangling procedure, making clusters with double heralding, quantum computer architecture.}\\
\cline{Quantum computing with optical nonlinearities}{kerr}\\
\cdes{Weak cross-Kerr nonlinearities, deterministic parity gate, Zeno gate.}

\section*{Introduction}

\noindent
A quantum computer is a machine that can perform certain calculations much faster than a classical computer by using the laws of quantum mechanics. Quantum computers do not exist yet, because it is extremely difficult to control quantum mechanical systems to the necessary degree. What is more, we do at this moment not know {\em which physical system} is the best suited for making a quantum computer (although we have some ideas). It is likely that a mature quantum information processing technology will use (among others) light, because photons are ideal carriers for quantum information. These notes are an expanded version of the five lectures I gave on the possibility of making a quantum computer using light, at the Summer School in Theoretical Physics in Durban, 14-24 January, 2007. There are quite a few proposals using light for quantum computing, and I can highlight only a few here. I will focus on photonic qubits, and leave out continuous variables completely\footnote{For a review on optical quantum computing with continuous variables, see Braunstein and Van Loock, {\it Rev. Mod. Phys.} {\bf 77}, 513 (2005).}. I assume that the reader is familiar with basic quantum mechanics and introductory quantum computing.

\section{Light and quantum information}\label{lec:light}

Simply put, a quantum computer works by storing information in physical carriers, which then undergo a series of unitary (quantum) evolutions and measurements. The information carrier is usually taken to be a {\em qubit}, a quantum system that consists of two addressable quantum states. Furthermore, the qubit can be put in arbitrary superposition states. The unitary evolutions on the qubits that make up the computation can be decomposed in single-qubit operations and two-qubit operations. Both types of operations or {\em gates} are necessary if the quantum computer is to outperform any classical computer.

\subsection{Photons as qubits}

We define the {\em computational basis states} of the qubit as some suitable set of states $|0\rangle$ and $|1\rangle$. An arbitrary single-qubit operation can take the form of a compound rotation parameterized by two angles $\theta$ and $\phi$:
\begin{eqnarray}
	|0\rangle &\rightarrow& \cos\theta\, |0\rangle + i e^{i\phi} \sin\theta\, |1\rangle , \cr
	|1\rangle &\rightarrow& i e^{i\phi} \sin\theta\, |0\rangle + \cos\theta\, |1\rangle .
	\label{eq:single-qubit}
\end{eqnarray}
This can be represented graphically in the {\em Bloch} or {\em Poincar\'e} sphere
\vskip-0.8cm
\begin{figure}[bh]
  \begin{center}
  \begin{psfrags}
    \psfrag{a}{$|0\rangle$}
    \psfrag{b}{$|1\rangle$}
    \psfrag{c}[r]{$|-\rangle = |0\rangle - |1\rangle$}
    \psfrag{d}{$|+\rangle = |0\rangle + |1\rangle$}
    \epsfig{file=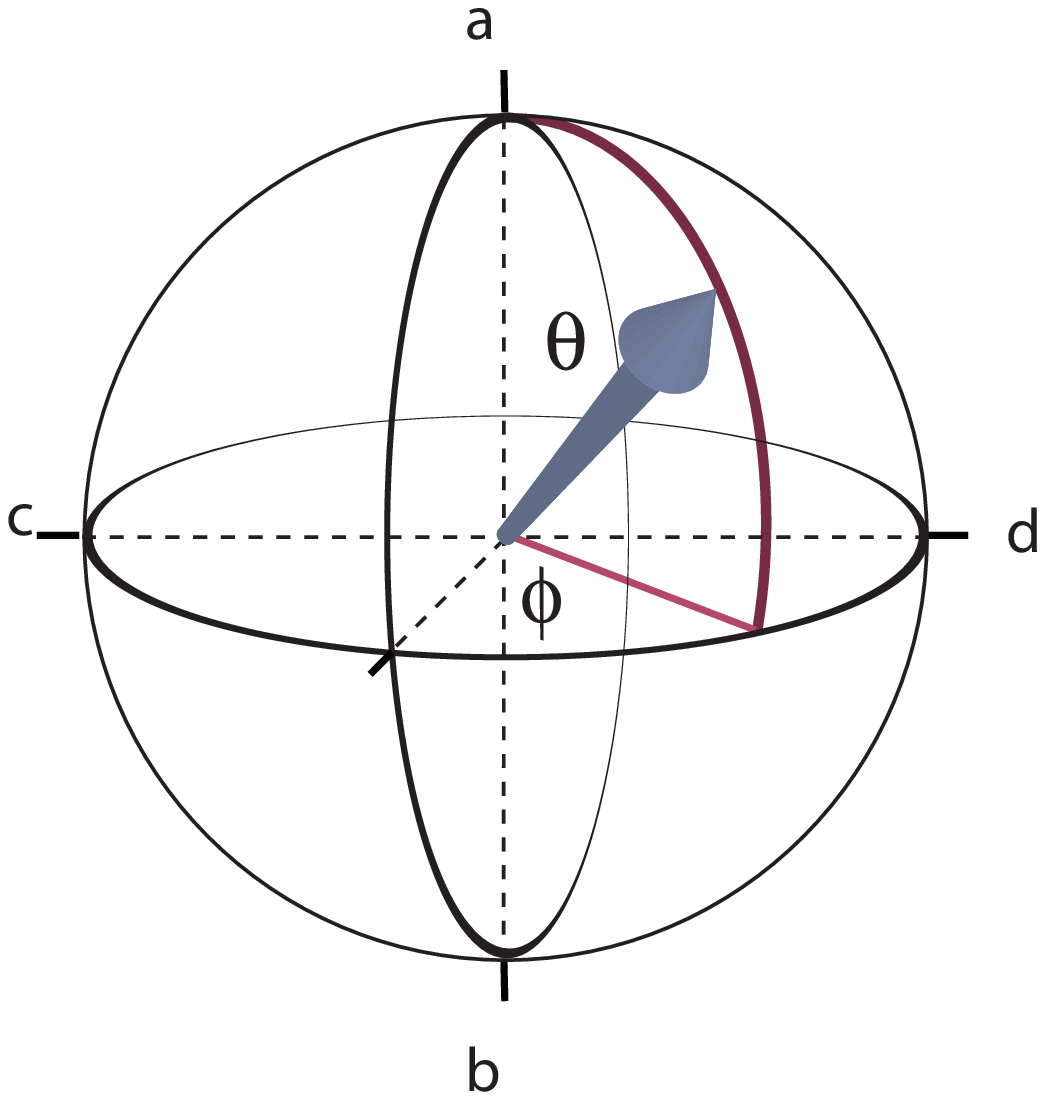,height=5.8cm}
  \end{psfrags}
  \end{center}
\end{figure}

What type of light can be used as a qubit? The smallest excitation of the electromagnetic field is the {\em photon}. We cannot construct a standard wave function for the photon, but we can identify the different degrees of freedom that we can use as a qubit: A photon can have the choice between {\em two spatially separated beams} (or modes), or it can have two distinct {\em polarizations} \cite{kok06}. These two representations are mathematically equivalent, as we will show below.

The emission and absorption of photons with momentum $k$ is described mathematically using {\em creation} and {\em annihilation} operators:
\begin{equation}
 \hat{a}(k) |n\rangle_k = \sqrt{n} |n-1\rangle_k \quad\text{and}\quad
 \hat{a}^{\dagger}(k) |n\rangle_k = \sqrt{n+1} |n+1\rangle_k \; .
\end{equation}
It is straightforward to show
that $\hat{n}(k) \equiv \hat{a}^{\dagger}(k) \hat{a}(k)$ is the number
operator $\hat{n}(k) |n\rangle_k = n |n\rangle_k$. The canonical commutation relations between $\hat{a}$
and $\hat{a}^{\dagger}$ are given by 
\begin{eqnarray}\label{eq:commutation}
 \left[\hat{a}(k),\hat{a}^{\dagger}(k')\right] &=& \delta(k-k'), \cr
 \left[\hat{a}(k),\hat{a}(k')\right] &=& \left[\hat{a}^{\dagger}(k),
 \hat{a}^{\dagger}(k')\right] = 0\; .
\end{eqnarray}
For the purposes of these notes, we use subscripts to distinguish the creation and annihilation operators for different modes, rather than the functional dependence on $k$.
In photon language, we can define the logical qubit states on two spatial modes $a$ and $b$ as:
\begin{eqnarray}
 |0\rangle_L &=& \hat{a}^{\dagger} |\vac\rangle = |1,0\rangle_{ab} \qquad\text{and} \cr 
 |1\rangle_L &=& \hat{b}^{\dagger} |\vac\rangle = |0,1\rangle_{ab},
\end{eqnarray}
where $|\vac\rangle$ is the vacuum state and the 0 and 1 denote the photon numbers in the respective modes. The polarization qubits are defined as
\begin{eqnarray}
 |0\rangle_L &=& \hat{a}_H^{\dagger} |\vac\rangle = |H\rangle \qquad\text{and}\quad \cr
 |1\rangle_L &=& \hat{a}_V^{\dagger} |\vac\rangle = |V\rangle .
\end{eqnarray}
Every state of the electromagnetic field can be written as a function of the creation operators acting on the vacuum state $|\vac\rangle$. A change in the state can therefore also be described by a change in the creation operators (essentially, this is the difference between the Schr\"odinger and Heisenberg picture). In fact, it is often easier to work out how a physical operation changes the creation and annihilation operators than how it changes an arbitrary state. This is what we will do here. The single-qubit operations on single photons in terms of the creation and annihilation operators consist of the following transformations:
\begin{enumerate}
	\item The {\bf phase shift} changes the phase of the electromagnetic field in a given mode:
	\begin{equation}\label{ps2}
 	\hat{a}^{\dagger}_{\rm out} = e^{i\phi\hat{a}_{\rm
 	in}^{\dagger}\hat{a}_{\rm in}}\, \hat{a}_{\rm in}^{\dagger}\,
 	e^{-i\phi\hat{a}_{\rm in}^{\dagger} \hat{a}_{\rm in}} =
 	e^{i\phi} \hat{a}_{\rm in}^{\dagger}\; , 
	\end{equation}
	with the interaction Hamiltonian $H_{\phi}=\phi\,\hat{a}_{\rm in}^{\dagger}\hat{a}_{\rm in}$ ($\hbar=1$). Physically, the phase shift can be implemented using a delay line or a transparent element with an index of refraction that is different from free space, or the optical fiber (or whatever medium the photons propagate through). In Eq.~(\ref{ps2}) we used the operator identity 
	\begin{equation}
	  e^{\alpha A} B\, e^{-\alpha A} = B + \alpha [A,B] + \frac{\alpha^2}{2!} [A,[A,B]] + \ldots , 
	\end{equation}  
	  where $A$ is Hermitian.
	\item The {\bf beam splitter} usually consists of a semi-reflective mirror: when light falls on this mirror, part will be reflected and part will be transmitted. Let the two incoming modes on either side of the beam splitter be denoted by $\hat{a}_{\rm in}$ and $\hat{b}_{\rm
  in}$, and the outgoing modes by $\hat{a}_{\rm out}$ and $\hat{b}_{\rm out}$. When we parameterize the probability amplitudes of these possibilities as $\cos\theta$ and $\sin\theta$, and the relative phase as $\varphi$, then the beam splitter yields an evolution in operator form
\begin{eqnarray}\label{2bs}
 \hat{a}_{\rm out}^{\dagger} &=& \cos\theta\,\hat{a}_{\rm in}^{\dagger} + 
 i e^{-i\varphi} \sin\theta\,\hat{b}_{\rm in}^{\dagger} \; , \cr
 \hat{b}_{\rm out}^{\dagger} &=& i e^{i\varphi}\sin\theta\,\hat{a}_{\rm
 in}^{\dagger} +  \cos\theta\,\hat{b}_{\rm in}^{\dagger} \; . 
\end{eqnarray}
In terms of the Hamiltonian evolution, we have 
\begin{equation}
	\hat{a}^{\dagger}_{\rm out} = e^{i H_{\rm BS}}\, \hat{a}^{\dagger}_{\rm in}\, e^{-i H_{\rm BS}}  \qquad\text{and}\qquad
	\hat{b}^{\dagger}_{\rm out} = e^{i H_{\rm BS}}\, \hat{b}^{\dagger}_{\rm in}\, e^{-i H_{\rm BS}} , 
	\label{eq:bsevo}
\end{equation}
where the `interaction Hamiltonian' $H_{\rm BS}$ is given by
\begin{equation}\label{su2plus}
 H_{\rm BS} = \theta e^{i\varphi} \hat{a}_{\rm  in}^{\dagger}\hat{b}_{\rm in}
 + \theta e^{-i\varphi} \hat{a}_{\rm in}\hat{b}_{\rm in}^{\dagger}\; .
\end{equation}
Mathematically, the two parameters $\theta$ and $\varphi$ represent the angles of a rotation about two orthogonal axes in the Poincar\'e sphere. The physical beam splitter can be described by any choice of $\theta$ and $\varphi$, where $\theta$ is a measure of the transmittivity, and $\varphi$ gives the phase shift due to the coating of the mirror. An additional phase shift may be necessary to describe the workings of the physical object correctly.
\end{enumerate}
This demonstrates that the beam splitter and the phase shift suffice to implement any single-qubit operation on a single photonic qubit. This case, where a single photon can be in two optical modes, is commonly called the {\em dual rail} representation, as opposed to the {\em single rail} representation where the qubit coincides with the occupation number of a single optical mode.

There are similar relations for transforming the polarization of a photon. Physically, the polarization is the spin degree of freedom of the photon. The photon is a spin-1 particle, but because it travels at the speed of light $c$, the longitudinal component is suppressed. We are left with two polarization states, which make an excellent qubit. The two important operations on polarization are:
\begin{enumerate}
	\item The {\bf polarization rotation} is physically implemented by quarter- and half-wave plates. We write $\hat{a}_{\rm in}\rightarrow\hat{a}_x$ and 
$\hat{b}_{\rm in}\rightarrow\hat{a}_y$ for some orthogonal set of 
coordinates $x$ and $y$ (i.e., $\langle x|y\rangle=0$). The parameters
$\theta$ and $\varphi$ are now angles of rotation:
\begin{eqnarray}\label{2polrot}
 \hat{a}_{x'}^{\dagger} &=& \cos\theta\,\hat{a}_x^{\dagger} + 
 i e^{-i\varphi} \sin\theta\,\hat{a}_y^{\dagger} \; , \cr
 \hat{a}_{y'}^{\dagger} &=& i e^{i\varphi} \sin\theta\,\hat{a}_x^{\dagger} + 
 \cos\theta\,\hat{a}_y^{\dagger} \; .
\end{eqnarray}
This evolution has the same Hamiltonian as the beam splitter, and it
formalizes the equivalence between polarization and two-mode logic.
  \item The {\bf polarizing beam splitter} (PBS) spatially separates modes with orthogonal polarization. If the PBS is cut to separate horizontal and
vertical polarization, the transformation of the incoming modes
($a_{\rm in}$ and $b_{\rm in}$) yields the following outgoing modes
($a_{\rm out}$ and $b_{\rm out}$):
\begin{eqnarray}\label{eq:pbs}
  \hat{a}_{{\rm in},H} \rightarrow \hat{a}_{{\rm out},H} &
  ~\text{and}~ & \hat{a}_{{\rm in},V} \rightarrow \hat{b}_{{\rm out},V}
  \cr 
  \hat{b}_{{\rm in},H} \rightarrow \hat{b}_{{\rm out},H} &
  ~\text{and}~ & \hat{b}_{{\rm in},V} \rightarrow \hat{a}_{{\rm out},V} .
\end{eqnarray}
Using quarter-wave plates and polarizers, we can also construct a PBS
for different polarization directions (e.g., $L$ and $R$), in which
case we make the substitution $H \leftrightarrow L$, $V
\leftrightarrow R$.
\end{enumerate}

\subsection{Interferometers}

When there are many optical modes $a_1$ to $a_N$, we need a compact description if we are to apply beam splitters, phase shifters and such to these optical modes. Equations (\ref{2bs}) and (\ref{2polrot}) can be written as a vector equation
\begin{equation}
	\begin{pmatrix}
	 \hat{a}_{\rm out}^{\dagger} \cr \hat{b}_{\rm out}^{\dagger}
	\end{pmatrix}
	=
	\begin{pmatrix}
	 \cos\theta & ie^{-i\varphi} \sin\theta \cr ie^{i\varphi} \sin\theta & \cos\theta
	\end{pmatrix}
	\begin{pmatrix}
	 \hat{a}_{\rm in}^{\dagger} \cr \hat{b}_{\rm in}^{\dagger}
	\end{pmatrix}.
	\label{eq:vectoreq}
\end{equation}
In general, when we have many optical modes we can collect their corresponding operators in a vector, and if $U$ is a unitary matrix, the multi-mode transformations become
\begin{equation}
	\hat{\vec{a}}_{\rm out}^{\dagger} = U\cdot\hat{\vec{a}}_{\rm in}^{\dagger} \quad\text{or}\quad \hat{a}_{j,\rm out}^{\dagger} = \sum_k U_{jk} \hat{a}_{k,\rm in}^{\dagger}\; ,
	\label{eq:bog}
\end{equation}
where $\hat{\vec{a}}_{\rm out} \equiv (\hat{a}_1,\ldots,\hat{a}_N)$. A successive application of beam splitters and phase shifters is therefore equivalent to a series of unitary matrices associated with these elements. It turns out that any $N\times N$ unitary matrix can be decomposed in terms of $2\times 2$ unitary matrices $T_{jk}$ of the form\footnote{To be precise, the $N\times N$ matrix is decomposed in terms of $T_{jk}\otimes \unity_{N-2}$, where $\unity_{N-2}$ is the $(N-2)\times(N-2)$ identity matrix.} in equation (\ref{eq:vectoreq}) \cite{reck}. Therefore, any arbitrary {\em interferometer} (in which $N$ optical modes interfere with each other) can be constructed from beam splitters, phase shifts, and polarization rotations. This is an extraordinarily powerful result, and we can use it to define a general interferometer as a unitary transformation $U$ on $N$ (spatial) modes, or an $N$-port. 

We should note one very important thing, though: Just because we can decompose $U$ into a series of ``single-qubit" operations defined above, it does not mean we can call this a quantum computer. Qubits should be well-defined physical systems that you can track through the computation. However, in an interferometer with $n$ input photons it is possible (and inevitable) that some of them will end up in the same mode. Since photons are indistinguishable particles (or at least they should be in this model), we cannot track the quantum information they carry. Also, we still haven't shown how to make two-qubit gates. This means that we have to work a bit harder to show we can make a quantum computer in this way.

\bigskip

\begin{exercise}
 Prove the relations in Eqs.~(\ref{ps2}) and (\ref{2bs}).
\end{exercise}

\begin{exercise}
 Show how to turn a qubit on two spatial modes into a polarization qubit.
\end{exercise}

\begin{exercise}
 Write down the interaction Hamiltonian and unitary matrix for a mirror.
\end{exercise}

\section{Two-qubit gates and the KLM scheme}\label{lec:klm}

While single-qubit operations on a photon are easy, two-qubit operations on two photons are very difficult. Consider the two-qubit gate that generates the following transformation:
\begin{equation}
	|H,H\rangle_{ab} \quad\rightarrow\quad \frac{1}{\sqrt{2}} \left( |H,H\rangle_{cd} + |V,V\rangle_{cd} \right)\; .
	\label{eq:bell}
\end{equation}
This is a perfectly sound quantum mechanical operation, and one that is often needed in a quantum computation. Let's see how we can implement this with photons and linear optical elements. In terms of the creation operators acting on the vacuum $|\vac\rangle$, this transformation can be written as
\begin{equation}
	\hat{a}_H^{\dagger} \hat{b}_H^{\dagger}|\vac\rangle \quad\rightarrow\quad \frac{1}{\sqrt{2}} \left( \hat{c}_H^{\dagger} \hat{d}_H^{\dagger} + \hat{c}_V^{\dagger} \hat{d}_V^{\dagger} \right)|\vac\rangle\; .
	\label{eq:bellops}
\end{equation}
Let's substitute the operator transformations for $\hat{a}_H^{\dagger}$ and $\hat{b}_H^{\dagger}$: 
\begin{equation}
	\hat{a}_H^{\dagger}\hat{b}_H^{\dagger} = \left( \sum_j U_{j1} \hat{c}_j^{\dagger} \right) \left( \sum_k U_{k2} \hat{d}_k^{\dagger} \right) = \sum_{jk} U_{j1} U_{k2} \hat{c}_k^{\dagger} \hat{d}_j^{\dagger} \; .
	\label{eq:notwoqubitgate}
\end{equation}
By construction, this is a separable expression. However, the state we wish to create is entangled (inseparable)! So we can never get an entangling (two-qubit) gate this way. Therefore we arrive at the conclusion that single-photon inputs, $N$-ports and final read-out is not sufficient to make a quantum computer!

\subsection{The KLM approach}

Clearly, we have to add something more. What about feed-forward? By making a measurement on part of the output of the $N$-port we may be able to reject or accept certain terms in a superposition, and effectively gain entanglement. This is the approach championed in the now famous ``KLM'' paper \cite{klm}, after the authors Knill, Laflamme, and Milburn. First, they construct an $N$-port with suitable input states, which upon the correct detection signature gives a two-qubit gate. Since the detection is a true quantum mechanical process, the outcome is unknown beforehand, and the gate succeeds only a fraction of the time. The gate destroys the qubits (and hence the quantum information) when it fails. If this gate was used directly in a computation, the overall success probability of the computation would decrease exponentially with the number of two-qubit gates, so something else is needed.

Second, KLM show how a probabilistic optical gate can be applied to two qubits without destroying them. This relies on a method developed by Gottesman and Chuang, called the {\em teleportation trick}, and it allows us to use a previously created entangled state to teleport gates into the quantum circuit \cite{gott}. I will describe this procedure in a different form later in these lecture notes. Before that, let's look at quantum gates in a bit more detail.

Three special single-qubit gates are the Pauli operators. In matrix notation (in the computational basis) these look like
\begin{equation}
	X = 
	\begin{pmatrix}
	 0 & 1 \cr 1 & 0
	\end{pmatrix},
	\quad
	Y = 
	\begin{pmatrix}
	 0 & -i \cr i & 0
	\end{pmatrix},
	\quad
	Z = 
	\begin{pmatrix}
	 1 & 0 \cr 0 & -1
	\end{pmatrix}.
	\label{eq:pauli}
\end{equation}
The $X$ operator is a bit flip, and the $Z$ operator is a phase flip. The $Y$ operator is a combination of $X$ and $Z$. Two very useful two-qubit gates are the controlled-$Z$, where a $Z$ operation is applied to the second qubit if the first qubit is in state $|1\rangle$, and the controlled-{\sc not}, where an $X$ operation (a bit flip) is applied to the second qubit depending on the first. In matrix notation, these gates look like
\begin{equation}
	U_{CZ} = 
	\begin{pmatrix}
	 1 & 0 & 0 & 0 \cr 0 & 1 & 0 & 0 \cr 0 & 0 & 1 & 0 \cr 0 & 0 & 0 & -1
	\end{pmatrix}
	\quad\text{and}\quad
	U_{CNOT} = 
	\begin{pmatrix}
	 1 & 0 & 0 & 0 \cr 0 & 1 & 0 & 0 \cr 0 & 0 & 0 & 1 \cr 0 & 0 & 1 & 0
	\end{pmatrix}.
	\label{eq:CZCNOT}
\end{equation}
These two entangling gates have a very special property: When we apply these transformations to a tensor product of two Pauli matrices we get again two Pauli matrices:
\begin{equation}
	U_{CZ}^{\dagger} (P_1 \otimes P_2) U_{CZ} = P_3 \otimes P_4\; ,
	\label{eq:clifford}
\end{equation}
and similarly for the {\sc cnot} gate. Operators with this property (of turning Pauli operators into Pauli operators) are members of the {\em Clifford} group. This is a very important symmetry in quantum information theory, as it forms the basis of quantum error correction.

\begin{figure}[t]
  \begin{center}
    \epsfig{file=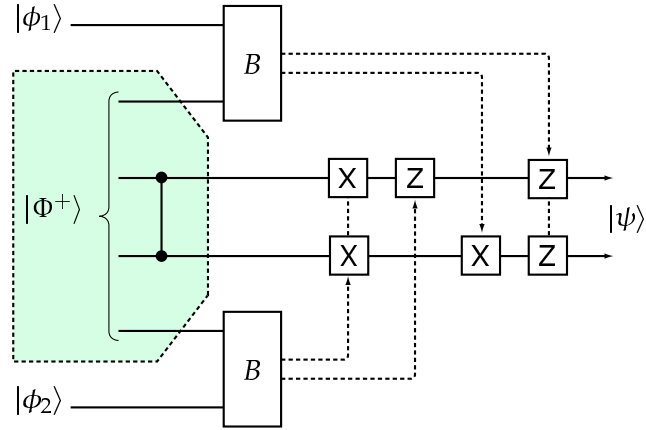}
  \end{center}
  \caption{\small The teleportation trick. The {\sc cz} operation is denoted by a vertical line, which connects to the two qubits with a solid dot. We teleport both qubits $|\phi_1\rangle$ and $|\phi_2\rangle$ ($B$ denotes the Bell measurement), and apply the {\sc cz} to the output qubits. Then we commute the {\sc cz} from the right to the left, through the corrective Pauli operations of the teleportation. The {\sc cz} operation can then be performed {\em off-line}, together with the preparation of the entanglement channel for teleportation (the green box).}
  \label{fig:tele}
\end{figure}

Suppose we wish to apply the {\sc cz} gate to two qubits $|\phi_1\rangle$ and $|\phi_2\rangle$ (see fig.~\ref{fig:tele}). We can teleport these states to new qubit systems and then apply the {\sc cz} gate to the teleported qubits. This in itself achieves not much, but we can now commute the {\sc cz} gate through the corrective single-qubit Pauli gates to make the {\sc cz} part of the entanglement channel in teleportation. The fact that the {\sc cz} operation is part of the  Clifford group now comes in handy: The commutation operation will {\em not} induce any new two-qubit gates. 

Knill, Laflamme, and Milburn \cite{klm} used this trick to create two-qubit gates for single-photon qubits. The complication here was that the Bell measurement essential to teleportation cannot be carried out deterministically on single photons. To this end, KLM designed a teleportation protocol that uses $2n$ additional photons and succeeds with a success probability $n/(n+1)$. Since we require two teleportation events, the success probability of the two-qubit gate is $[n/(n+1)]^2$. Failure of the gate amounts to a measurement in the computational basis, which is easy to protect against with standard error correction (i.e., parity codes). 

This may all seem a bit overwhelming, and the reader will be pleased to hear that several simplifications of this scheme have been proposed. In the next part of this lecture I will describe two very simple optical operations that can be used to create all the entanglement we need.

\subsection{Two-photon interference}

Quantum computing with photons and linear optical elements relies critically on two-photon interference, with or without polarization. In this section, I will first describe the quintessential two-photon Hong-Ou-Mandel effect. After that, I will extend it to the case of polarized photons. 

\begin{figure}[t]
  \begin{center}
    \epsfig{file=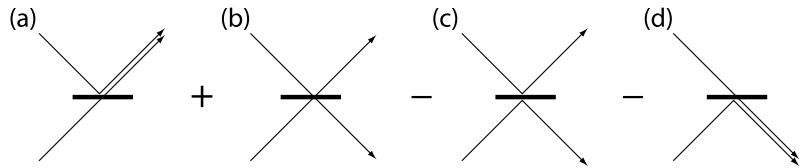,width=10.5cm}
  \end{center}
  \caption{\small The four amplitudes in the two-photon interference experiment by Hong, Ou, and Mandel. Components (b) and (c) always have opposite sign by virtue of unitarity of the beam splitter, and cancel.}
  \label{fig:hom}
\end{figure}

The Hong-Ou-Mandel (HOM) effect \cite{hom} occurs when two {\em identical} photons (the same polarization, the same frequency, and the same spatio-temporal profile) each enter an input port of a 50:50 beam splitter (see Fig.\  \ref{fig:hom}). Mathematically, this can be condensed to the following: Since the photons are identical, we can suppress all the spatio-temporal, frequency, and polarization information in the creation operator, and write the input state as $|1,1\rangle_{ab} = \hat{a}^{\dagger} \hat{b}^{\dagger} |\vac\rangle$ on the two input modes $a$ and $b$. The 50:50 beam splitter is characterized by the transformation
\begin{equation}
	\hat{a} \rightarrow \frac{\hat{c}+\hat{d}}{\sqrt{2}} \quad\text{and}\quad
  \hat{b} \rightarrow \frac{\hat{c}-\hat{d}}{\sqrt{2}}\; .
	\label{eq:5050}
\end{equation} 
Classically, when the photons enter the beam splitter, each will independently choose whether it will exit in mode $c$ or $d$. As a result, we expect the photons half of the time to come out in the same output (both in $c$ or both in $d$), and half of the time they should come out in different output modes (one in mode $c$ and one in mode $d$). However, quantum mechanically we get something different.

When we substitute the beam splitter transformation rules into the input state $|1,1\rangle_{ab}$, we obtain
\begin{eqnarray}
	|1,1\rangle_{ab} &=& \hat{a}^{\dagger}\hat{b}^{\dagger} |\vac\rangle \rightarrow \frac{1}{2} \left( \hat{c}^{\dagger} + \hat{d}^{\dagger} \right) \left( \hat{c}^{\dagger}-\hat{d}^{\dagger} \right) |\vac\rangle = \frac{1}{2} \left( \hat{c}^{\dagger 2} - \hat{d}^{\dagger 2} \right) |\vac\rangle \cr &=&  \frac{|2,0\rangle_{cd} - |0,2\rangle_{cd}}{\sqrt{2}}\; .
	\label{eq:hom}
\end{eqnarray}
We see that the $|1,1\rangle_{cd}$ term in the output modes of the beam splitter is suppressed. This is the HOM effect, and the absence of coincidence counts in such an interference experiment is called the HOM dip. When the input photons are distinguishable (for example if they have different frequencies, or if they arrive at different times at the beam splitter), the dip disappears, and we see the $|1,1\rangle_{cd}$ component in the superposition: The photons behave as classical particles. It is therefore extremely important in such experiments that the photons are truly indistinguishable. This is one of the hardest requirements to meet in linear optical quantum computing.

Another way to see how the HOM effect works is to write down all the different possibilities in which the photons can travel through the beam splitter (see Fig.\  \ref{fig:hom}). The output state of (b) and (c) are indistinguishable, so we do not know whether both photons were transmitted or reflected. Moreover, the beam splitter does not retain a memory how the photons interacted at its surface. Therefore, we have to sum the two possibilities coherently. Unitarity of the beam splitter ensures that the relative phase is $-1$, and the two processes cancel. We can run this experiment backwards as well, because a unitary evolution is reversible. The sources then become detectors, and vice versa. It is then easy to see that a coincidence count after a 50:50 beam splitter projects onto the state $|2,0\rangle - |0,2\rangle$ of the input modes. 

\begin{figure}[t]
  \begin{center}
    \epsfig{file=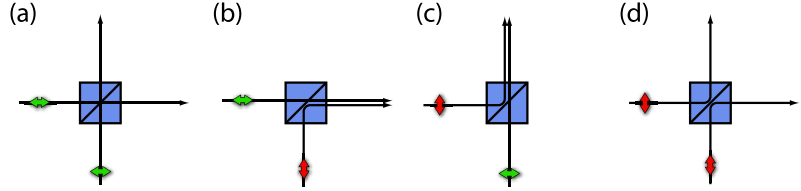,width=10.5cm}
  \end{center}
  \caption{\small The effect of a polarizing beam splitter on two input photons. The red and green arrows denote vertical and horizontal polarization, respectively.}
  \label{fig:fusion}
\end{figure}

The HOM effect is the corner stone of KLM-type optical quantum computing. When the qubit is a dual-rail single photon, every two-qubit gate is based on this effect. However, it may be sometimes more convenient to use polarization qubits. Can we construct a similar two-qubit interferometer? The answer is yes: Assume that we have two photons impinging on a polarizing beam splitter. In Fig.\  \ref{fig:fusion} you can see that the action of this device looks very similar, except that there is no cancellation. When we erase polarization information in the output modes by $45^{\circ}$ rotations and perform single-photon detection, we can construct so-called {\em fusion gates}. These turn out to be extremely useful for optical quantum computing.

There are two types of fusion gates, aptly named type I and type II \cite{browne}. In type I, only one of the output ports is detected, while in type II both output ports are detected. Because of this detection, if we want to create entanglement we cannot start with single photons. The basic building block that is to be used with fusion gates is a Bell state, e.g., $|H,V\rangle + |V,H\rangle$. It is not easy to make these states on demand, but there are quite a few experimental efforts underway to create them with micro-pillar structures. I will first describe the precise workings of both fusion gates, and then I will show how they can be used to make large sets of entangled qubits.

\begin{figure}[t]
  \begin{center}
       \epsfig{file=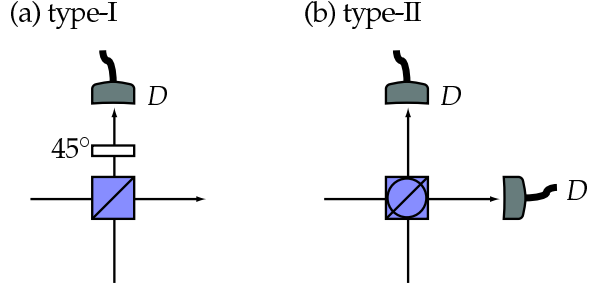}
  \end{center}
 \caption{\small Two types of fusion operators. (a) the type-I fusion operator
  employs a polarization beam splitter (PBS1) followed by the detection
  $D$ of a single output mode in the 45$^{\circ}$ rotated polarization
  basis. This operation determines the parity of the input mode with
  probability 1/2. (b) the type-II fusion operator uses a diagonal
  polarization beam splitter (PBS2), detects both output modes, and
  projects the input state onto a maximally entangled Bell state with
  probability 1/2.}  
 \label{fig:fusion2}
\end{figure}

The type-I fusion gate is a polarizing beam splitter cut for horizontal and vertical polarization, and one of the output modes (say, $d$) has a $45^{\circ}$ polarization rotation, followed by photo-detection in the $\{H,V\}$ basis. Let's assume that the input modes $a$ and $b$ are entangled with some other modes, such that the most general input state can be written as $(f_1 \hat{a}_H^{\dagger} + f_2 \hat{a}_V^{\dagger})(f_3 \hat{b}_H^{\dagger} + f_4 \hat{b}_V^{\dagger})|\vac\rangle$. Here, the $f_k$ are arbitrary functions of creation operators on other optical modes. When we substitute the transformation of the polarizing beam splitter and the polarization rotation, we obtain the following operator
\begin{equation}\nonumber
	\frac{f_1 f_3}{\sqrt{2}} \hat{c}^{\dagger}_H \left( \hat{d}^{\dagger}_H +  \hat{d}^{\dagger}_V \right) + f_1 f_4 \hat{c}^{\dagger}_H \hat{c}^{\dagger}_V + \frac{f_2 f_3}{2} \left( \hat{d}^{\dagger 2}_H - \hat{d}^{\dagger 2}_V \right) + \frac{f_2 f_4}{\sqrt{2}} \hat{c}^{\dagger}_V \left( \hat{d}^{\dagger}_H -  \hat{d}^{\dagger}_V \right) .
\end{equation}
After post-selecting the output state of all the modes (including the support of the $f_k$) on the detector outcome $(d_H,d_V)$, we have
\begin{eqnarray}\label{eq:type1}
 (0,0) &:\quad& |\psi_{\rm out}\rangle = f_1 f_2 \hat{c}^{\dagger}_H \hat{c}^{\dagger}_V |\vac\rangle \cr
 (2,0) ~\text{or}~ (0,2) &:\quad& |\psi_{\rm out}\rangle = \frac{f_2 f_3}{2} |\vac\rangle \cr
 (1,0) &:\quad& |\psi_{\rm out}\rangle = \frac{1}{\sqrt{2}} \left( f_1 f_3 \hat{c}^{\dagger}_H + f_2 f_4 \hat{c}^{\dagger}_V \right)|\vac\rangle \cr
 (0,1) &:\quad& |\psi_{\rm out}\rangle = \frac{1}{\sqrt{2}} \left( f_1 f_3 \hat{c}^{\dagger}_H - f_2 f_4 \hat{c}^{\dagger}_V \right)|\vac\rangle \; .
\end{eqnarray}
In the case where we find a single photon in mode $d$ (vertical or horizontal), it is easy to see that we create entanglement. In particular (and with abusive notation), suppose that $f_1 = \hat{a}_H^{\dagger}$, $f_2 = \hat{a}_V^{\dagger}$, $f_3 = \hat{b}_H^{\dagger}$, and $f_4 = \hat{b}_V^{\dagger}$. The arrival of a horizontal photon in mode $d$ then signals the output state $(\hat{a}_H^{\dagger} \hat{b}_H^{\dagger} \hat{c}_H^{\dagger} + \hat{a}_V^{\dagger} \hat{b}_V^{\dagger} \hat{c}_V^{\dagger}) |\vac\rangle$, that is, a three-photon GHZ state.

The type-II fusion operator in fig.~\ref{fig:fusion2}b works in a very similar way to the type-I fusion gate, except now the polarizing beam splitter is cut to diagonal polarization $|H\rangle\pm |V\rangle$, and both output ports are detected in the $\{H,V\}$ basis. When we again substitute the transformation of the polarizing beam splitter and the polarization rotation, we obtain the operator
\begin{eqnarray}\nonumber
	&& (f_1 + f_2)(f_3 - f_4) \left( \hat{c}_H^{\dagger 2} - \hat{c}_V^{\dagger 2} \right) + 
	(f_1 - f_2)(f_3 + f_4) \left( \hat{d}_H^{\dagger 2} - \hat{d}_V^{\dagger 2} \right) \cr && 
	+ 2(f_1 f_3 + f_2 f_4) \hat{c}_H^{\dagger} \hat{d}_H^{\dagger} + 2(f_1 f_4 + f_2 f_3) 
	\hat{c}_H^{\dagger} \hat{d}_V^{\dagger} + 2(f_1 f_4 + f_2 f_3) \hat{c}_V^{\dagger} 
	\hat{d}_H^{\dagger}\cr && + 2(f_1 f_3 + f_2 f_4) \hat{c}_V^{\dagger} \hat{d}_V^{\dagger}.
\end{eqnarray}
Depending on the photon detection signature $(c,d)$, we have the output state
\begin{eqnarray}\label{eq:type2}
 (2H,0) ~\text{or}~ (2V,0) &:\quad& |\psi_{\rm out}\rangle = (f_1 + f_2)(f_3 - f_4) |\vac\rangle \cr
 (0,2H) ~\text{or}~ (0,2V) &:\quad& |\psi_{\rm out}\rangle = (f_1 - f_2)(f_3 + f_4) |\vac\rangle \cr
 (H,H) ~\text{or}~ (V,V) &:\quad& |\psi_{\rm out}\rangle = (f_1 f_3 + f_2 f_4) |\vac\rangle \cr
 (H,V) ~\text{or}~ (V,H) &:\quad& |\psi_{\rm out}\rangle = (f_1 f_4 + f_2 f_3) |\vac\rangle\; .
\end{eqnarray}
Clearly, when we find one photon in each output port, the type-II fusion gate is an entangling gate. It can be interpreted as a {\em parity} measurement as follows: Suppose that the functions $f_k$ are not operators, but quantum state amplitudes of the two input modes $a$ and $b$ instead. The input state is then given by $f_1 f_3 |H,H\rangle + f_2 f_4 |V,V\rangle + f_1 f_4 |H,V\rangle + f_2 f_3 |V,H\rangle$. Clearly, finding two photons with {\em the same} polarization in the output modes will project onto the {\em even parity} component of the input state, which finding two photons with {\em different} polarization will project onto the {\em odd parity} component.

The fusion gates are not your regular two-qubit gates, because you can't put a separable state in and get an entangled state out. In fact, there is no output in the type-II fusion gate at all. So how can we do quantum computing with this? The answer was given by Browne and Rudolph \cite{browne}, and it involves a whole new approach to quantum computing. It is known as the ``one-way model" of quantum computing, ``cluster state quantum computing", of the more generic ``measurement-based quantum computing". I will discuss the basic principles of this approach in the next section.

\bigskip
 
\begin{exercise}
 Can you make {\em any} entanglement with single photons and $N$-ports?
\end{exercise}

\begin{exercise}
 Show that the Hadamard, {\sc cz} and {\sc cnot} operators are members of the Clifford group.
\end{exercise}

\begin{exercise}
 Calculate the success probability of the type-I fusion gate.
\end{exercise}

\begin{exercise}
 Verify the relations (\ref{eq:type1}) and (\ref{eq:type2}). Convince yourself that the type-II fusion gate is a parity projection.
\end{exercise}

\section{Cluster states}\label{lec:cluster}

Cluster states were introduced by Raussendorf and Briegel \cite{rb}, and form an alternative approach to quantum computing. The heart of this architecture is to create a large entangled state as a resource. The computation then proceeds as a series of (parallel) {\em single-qubit measurements}. Since all the entanglement is produced ``off-line'', this is a particularly powerful approach for single-photon quantum computing. 

\subsection{From circuits to clusters}

Before we introduce cluster states, we must cast arbitrary single-qubit rotations into arbitrary rotations around the $Z$ axis and Hadamard operations $H$. Using $H^2 = \unity$ and $X = H Z H$, any arbitrary rotation $\vec{\theta}$ can be decomposed into three Euler angles $\alpha$, $\beta$, and $\gamma$:
\begin{equation}
	R(\vec{\theta}) = Z(\gamma) X(\beta) Z(\alpha) = H\, H Z(\gamma)\, H Z(\beta)\, H Z(\alpha)
	\label{eq:euler}
\end{equation}
where $Z(\alpha) \equiv \exp(i\alpha Z/2)$ and $X(\beta) \equiv \exp(i\beta X/2)$. In circuit language, this becomes
\medskip
\[
\Qcircuit @C=0.7em @R=.7em @!R {
& \qw & \gate{H Z(\alpha)} & \qw & \gate{H Z(\beta)} & \qw &\gate{H Z(\gamma)} & \qw & \gate{H} & \qw
}
\]
and I will now show how $H Z(\alpha)$ can be implemented via a single-qubit measurement.

Consider the following circuit diagram of single-qubit teleportation:
\medskip
\[
\Qcircuit @C=0.7em @R=.7em @!R {
& \lstick{\phantom{U} \ket{\psi}}  & \ctrl{1} & \qw & \qw & \gate{H} & \qw &\meter & \rstick{``m"} \\
& \lstick{\ket{0}} & \targ & \qw&\qw&\qw& \qw & \qw & \qw & \rstick{Z^m\ket{\psi}}
}
\] 
The measurement is in the computational basis, and the outcome ``$m$'' takes the value 0 or 1. Depending on this value, we apply a Pauli $Z$ operation to the teleported qubit.

The next step is to translate the {\sc cnot} gate into the {\sc cz} gate. This procedure incurs two Hadamard gates, which are absorbed into the state of the ancilla ($|0\rangle \to |+\rangle$) and the teleported qubit.
\medskip
\[
\Qcircuit @C=0.7em @R=.7em @!R {
& \lstick{\phantom{U} \ket{\psi}}  & \ctrl{1} & \qw & \qw & \gate{H} & \qw &\meter & \rstick{``m"} \\
& \lstick{\ket{+}} & \ctrl{-1} & \qw&\qw&\qw& \qw & \qw & \qw & \rstick{H Z^m\ket{\psi} = X^m H\ket{\psi}}
}
\] 
\medskip

A single-qubit rotation around the $Z$ axis to the input qubit $|\psi\rangle$ can be written as:
\medskip
\[
\Qcircuit @C=0.7em @R=.7em @!R {
& \lstick{\phantom{U} \ket{\psi}}  & \gate{Z(\alpha)} & \ctrl{1} & \gate{H} & \qw &\meter & \rstick{``m"} \\
& \lstick{\ket{+}} & \qw & \ctrl{-1} & \qw&\qw& \qw &\qw & \rstick{X^m H Z(\alpha)\ket{\psi}}
}
\]
\medskip

The rotation around the $z$-axis commutes with the {\sc cz} operation (they are both diagonal in the computational basis), and we can write:
\medskip
\[
\Qcircuit @C=0.7em @R=.7em @!R {
& \lstick{\phantom{U} \ket{\psi}}  & \ctrl{1} & \qw &\gate{H Z(\alpha)} & \meter & \rstick{``m"} \\
& \lstick{\ket{+}} & \ctrl{-1} & \qw &  \qw& \qw & \qw & \rstick{X^m H Z(\alpha)\ket{\psi}}
}
\]
\medskip

We can reinterpret this diagram as an entangled state $|\Psi\rangle = CZ |\psi,+\rangle$, followed by a single-qubit measurement on the first qubit that performs the single-qubit gate $H Z(\alpha)$ on the state $|\psi\rangle$. The precise measurement basis $A(\alpha)$ is determined by
\begin{equation}
 \langle\psi| Z(-\alpha) H\, Z\, H Z(\alpha) |\psi\rangle = \langle\psi|Z(-\alpha) X Z(\alpha) |\psi\rangle = \langle\psi| A(\alpha) |\psi\rangle\; .
\end{equation}
This corresponds to a measurement along an axis in the equatorial plane of the Bloch sphere.

In order to implement an arbitrary rotation $R(\vec{\theta})$, this procedure must be concatenated three times
\medskip
\[
\Qcircuit @C=0.7em @R=.7em @!R {
& \lstick{\phantom{U} \ket{\psi}}  & \ctrl{1}  & \qw       &  \qw       &  \qw       & \gate{H Z(\alpha)} & \meter & \rstick{``k"} \\
& \lstick{\ket{+}}                 & \ctrl{-1} & \ctrl{1}  &  \qw       &  \qw       & \gate{H Z(\beta)} & \meter & \rstick{``l"} \\
& \lstick{\ket{+}}                 & \qw       & \ctrl{-1} &  \ctrl{1}  &  \qw       & \gate{H Z(\gamma)} & \meter & \rstick{``m"} \\
& \lstick{\ket{+}}                 & \qw      & \qw       &  \ctrl{-1} &  \qw       & \qw  & \qw & \qw & \rstick{|\psi_{\rm out}\rangle} \\ &
}
\]
with $|\psi_{\rm out}\rangle = (X^m H Z(\gamma))\, (X^l H Z(\beta))\, (X^k H Z(\alpha))\ket{\psi}$. However, the operators $X^k$, $X^l$, and $X^m$ depend on the measurement outcomes, and we should try to get rid of them by commuting them through the Pauli gates and Hadamards. We can again use the relations $ZX= -XZ$ to show that 
\begin{equation}
	Z(\beta) X = \sum_{n=0}^{\infty} \left( \frac{i\beta}{2}\right)^n \frac{Z^n}{n!} \, X = \sum_{n=0}^{\infty} \left( \frac{-i\beta}{2}\right)^n \frac{X Z^n}{n!} = X Z(-\beta)\; .
	\label{eq:rotcom}
\end{equation} 
This therefore gives rise to an adjustment of the measurement bases depending on the previous measurement outcomes, and it results in a definite temporal direction in the computation (see Exercise~\ref{ex:oneway} at the end of this lecture). Hence the name ``one-way model" of quantum computing. (Remember that all the elements in the traditional circuit model are unitary operators, and therefore reversible.) 

\subsection{Universal cluster states}

Now that we have constructed arbitrary single-qubit operations, we do not need to start our circuit with the input state $|\psi\rangle$, but we can start with another $|+\rangle$ state and implement the first single-qubit rotation to obtain the required input state $|\psi\rangle$. We can then write the evolution of a single qubit graphically as a string of ancilla qubits in state $|+\rangle$ (circles), connected via $CZ$ operations (edges):
\begin{figure}[h]
  \begin{center}
       \epsfig{file=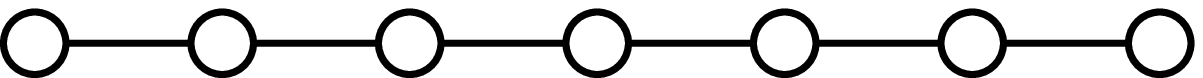, width=5cm}
  \end{center}
\end{figure}
\vskip -0.3cm

Multi-qubit evolution is then represented as a collection of such strings. The strings can be bridged vertically by edges, which in turn induce two-qubit operations: 
\vskip -0.3cm
\begin{figure}[h]
  \begin{center}
       \epsfig{file=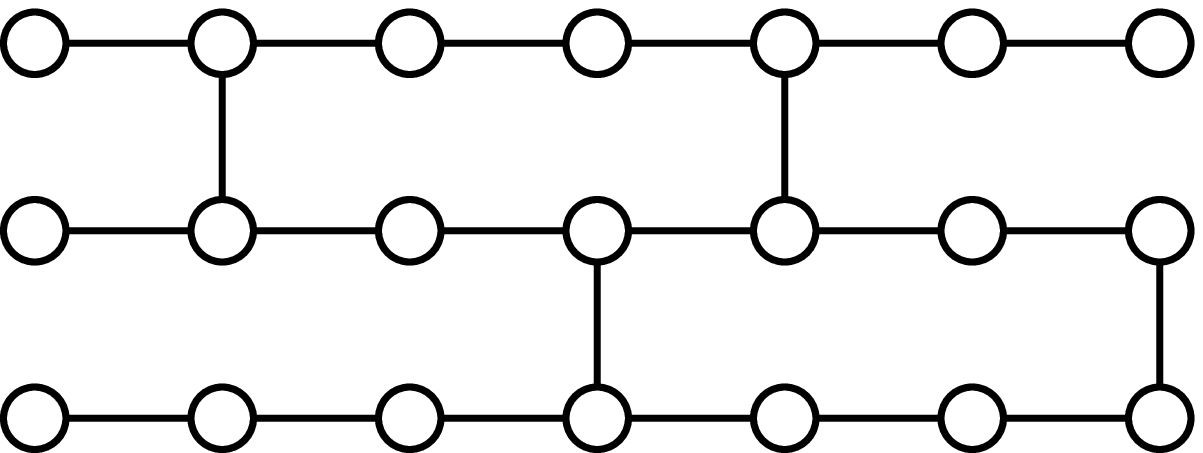, width=5cm}
  \end{center}
\end{figure}
\vskip -0.3cm

\noindent
When all the nearest-neighbour connections are established, and the qubits form an entangled grid, {\em any} quantum circuit can be realized if the cluster state is large enough. Such a state is called a {\em universal cluster state}.

To show that a vertical bridge induces a {\sc cz} gate, consider the following sequence of measurements and entangling operations: 
\begin{figure}[h]
  \begin{center}
       \epsfig{file=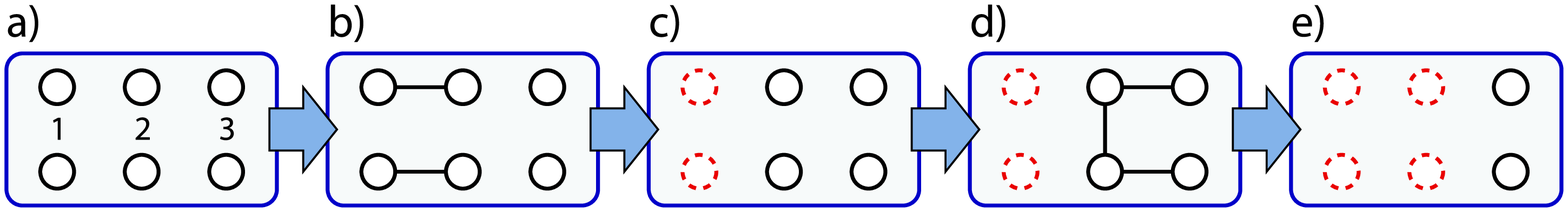, width=10.5cm}
  \end{center}
\end{figure}
\vskip -0.3cm

\noindent
We start with two rows of three qubits without any entanglement (a). The two rows will form the two qubits we wish to apply the {\sc cz} gate to. We then entangle qubit 1 with qubit 2 in each row (b) and measure qubits 1 (c). After the first two measurements, the states of qubits 1 are transferred to qubits 2. We can then apply the {\sc cz} gate to the two qubits, as well as the two {\sc cz} gates that connect qubits 2 with qubits 3 (d). Finally, we measure qubits 2 and transfer the quantum information to the output qubits 3 (e). The point here is that we can apply the {\sc cz} gate as if we are using it in the circuit model. However, because the measurements and the {\sc cz} operations in this sequence commute, we could have created all the entanglement at the start. Hence the vertical {\sc cz} gates are suitable as two-qubit gates in cluster state quantum computing. 

An often heard objection to cluster state quantum computing is that it seems to be very wasteful with entanglement. Instead of having to create entanglement for every two-qubit gate in an $N$-qubit computation, we seem to need at least $N$ entangling operations for {\em every} clock cycle! However, this is far too pessimistic. There is an enormous redundancy in a cluster state that is translated straight from the circuit model, as we did above. First of all, we can perform all single-qubit operations in the Clifford group {\em before} the computation starts: These operations also correspond to measurements, but their outcome does not affect any subsequent choice of measurement basis and can therefore be carried out at any stage. Also, these measurements will turn cluster states into smaller cluster states. As a result, we can calculate the effect of most Clifford operations and create a minimal cluster state that is in fact much smaller than what we found in the translation from the circuit model. Since most of the error correction in the quantum computation involves  Clifford operations, this is a huge saving. Secondly, we do not have to create the complete cluster state for the computation all at once. We can create a cluster with a relatively shallow depth, and keep adding qubits to the right as we measure qubits on the left. That way, we make our cluster ``just in time". The fusion gates can be used to create these cluster states with a moderate overhead per qubit, as I will show in the next section.

\subsection{Making cluster states with fusion gates}

In order to show that we can make cluster states with fusion gates we need two things. First, I will give a slightly unconventional description of the {\sc cz} gate, which allows us to give a formal description of a cluster state. Second, I will rewrite the action of the fusion gate in bracket notation.

A cluster state is a collection of qubits initially in the $|+\rangle$ state, with {\sc cz} operations applied to a set of qubit pairs. Remember that the {\sc cz} gate is defined as a $Z$ operation on the target\footnote{It so happens that the {\sc cz} gate is symmetric, so it does not really matter which qubit you call the control and which the target.} qubit if the state of the control qubit is $|1\rangle$. We can write this as
\begin{equation}
	U_{CZ}|+,\psi\rangle_{12} = |0\rangle_1 |\psi\rangle_2 + |1\rangle_1 \left( Z_2 |\psi\rangle_2 \right) ,
	\label{eq:weirdcz}
\end{equation}
where we ignored the overall normalization factor $1/\sqrt{2}$. We can do this, because all the terms have the same absolute value of the amplitude, and we are interested only in the relative phases. Now let's see what happens when there are multiple qubits in the cluster state. Qubit 1 may then have edges with multiple qubits:
\begin{equation}
	{\mathcal{U}}_{CZ}|+,\psi\rangle_{1..N} = |0\rangle_1 |\psi\rangle_{2..N} + |1\rangle_1 \left( Z_{j_1}\ldots Z_{j_k} |\psi\rangle_{2..N} \right) ,
	\label{eq:weirdcluster}
\end{equation}
where $Z_j$ is the Pauli $Z$ operator on qubit $j$, and ${\mathcal{U}}_{CZ}$ is a series of {\sc cz} gates on qubit 1 and its neighbors. To be accurate, we should write this as
\begin{equation}
	{\mathcal{U}}_{CZ}|+,\psi\rangle_{1..N} = |0\rangle_1 |\psi\rangle_{2..N} + |1\rangle_1  \prod_{j\in n(1)} {\mathcal{Z}}_j |\psi\rangle_{2..N} .
	\label{eq:halfcluster}
\end{equation}
Here, we defined ${\mathcal{Z}}_j = \unity_2\otimes\ldots Z_j\otimes\unity_{j+1}\ldots\otimes \unity_N$, and the neighborhood $n(1)$ is the set of qubits that are connected to qubit 1 via a {\sc cz} operation. Now suppose that we have two separate cluster states that we wish to fuse into one. We can write the separate states as
\begin{equation}
	\left( |0\rangle_1 |\psi\rangle_a + |1\rangle_1  \prod_{j\in n(1)} {\mathcal{Z}}_j |\psi\rangle_a \right) \otimes \left( |0\rangle_2 |\phi\rangle_b + |1\rangle_2  \prod_{j\in n(2)} {\mathcal{Z}}_j |\phi\rangle_b \right) .
	\label{eq:2clusters}
\end{equation}
So qubits 1 and 2 are connected to two different clusters states $|\psi\rangle$ and $|\phi\rangle$ on qubit sets $a$ and $b$, respectively. The neighbourhoods $n(1)$ and $n(2)$ have their support in these respective qubit sets.

We want to apply the type-I fusion gate to qubits 1 and 2, but before we can do this, we should write the action of the fusion gate in a more convenient form. From the last two lines of Eq.~(\ref{eq:type1}) we see that a single photon in either $d_H$ or $d_V$ heralds success, so let's assume we detect one photon in $d_H$. How does that transform the input to the output? In the discussion leading up to Eq.~(\ref{eq:type1}) we assumed that the state was given by 
\begin{equation}
	\left( f_1 f_3\, \hat{a}_H^{\dagger} \hat{b}_H^{\dagger} + f_1 f_4\, \hat{a}_H^{\dagger} \hat{b}_V^{\dagger} + f_2 f_3\, \hat{a}_V^{\dagger} \hat{b}_H^{\dagger} + f_2 f_4\, \hat{a}_V^{\dagger} \hat{b}_V^{\dagger} \right) |\vac\rangle .
	\label{eq:inputfusion}
\end{equation}
The fusion gate turns this into
\begin{equation}
	\left( f_1 f_3\, \hat{c}_H^{\dagger} + f_2 f_4\, \hat{c}_V^{\dagger} \right) |\vac\rangle ,
	\label{eq:outputfusion}
\end{equation}
and we can therefore deduce that only the $|H,H\rangle$ and $|V,V\rangle$ components survive. Moreover, the operator $\hat{c}_j^{\dagger}$ creates a photon with the same polarization as the ones that have just been detected: This can be written in bracket notation as
\begin{equation}
	U_{\rm type\, I}^{(H)} = |H\rangle \langle H,H| + |V\rangle \langle V,V|\; .
	\label{eq:brackettype1}
\end{equation}
A similar expression can be deduced for the case where a vertically polarized photon is detected.

To show that the type-I fusion gate can connect two cluster states, let's write $U_{\rm type\, I}$ in the computational basis: $U_{\rm type\, I}^{(0)} = |0\rangle_3 \,_{12}\langle 0,0| + |1\rangle_3 \,_{12}\langle 1,1|$. The fusion gate is now applied to qubit 1 and 2 in Eq.~(\ref{eq:2clusters}). This gives
\begin{eqnarray}
	|\psi_{\rm out}\rangle &=& |0\rangle_3 |\psi,\phi\rangle_{ab} + |1\rangle_3 \prod_{j\in n(1)}\prod_{k\in n(2)} {\mathcal{Z}}_j {\mathcal{Z}}_k |\psi,\phi\rangle_{ab} \cr
	&\equiv& |0\rangle_3 |\Psi\rangle_c + |1\rangle_3 \prod_{l\in n(1) \cup n(2)} {\mathcal{Z}}_l |\Psi\rangle_c ,
	\label{eq:merged}
\end{eqnarray}
where we defined $|\psi,\phi\rangle \equiv |\Psi\rangle$ and $c$ is the union of the two qubit sets $a$ and $b$. Note that Eq.~(\ref{eq:merged}) is again of the form in Eq.~(\ref{eq:halfcluster}), and is therefore another cluster state. If we had found measurement outcome 1 in the fusion gate, the same cluster state is created, up to a local $Z$ operation. This shows that we can use type-I fusion gates to create clusters states. The same is true for type-II gates.

Of course, the fusion gates are probabilistic, and half of the time the gate fails. It turns out that the type-II gate is better behaved than the type-I when it fails, and we should therefore aim to create the cluster states with type-II gates. However, we cannot just take the cluster state we want to expand and a single bell state and apply the type-II fusion, because in every successful gate we necessarily lose two photons through detection. We therefore need the type-I gate to create larger (but still small) cluster states, and use the type-II gate to add these to the cluster. How large should the mini-clusters be? 

Suppose we have a (linear) cluster of size $N$, and we want to add a mini-cluster of size $m$. The success probability of the fusion gate is $p$, and upon failure we need to detect one extra qubit to return the large multi-qubit state to a cluster state. In order for the cluster to grow we need to obey the following bound:
\begin{equation}
	p(N+m-2) + (1-p)(N-2) > N \quad\text{or}\quad m > \frac{2}{p}.
	\label{eq:bound}
\end{equation}
Therefore, even if the success probability is very small (for instance because of detection inefficiencies), we can still choose the size of our mini-clusters $m$ such that we can efficiently grow large cluster states. However, the larger $m$ is, the more the average cost of adding a qubit to the cluster, so we want $p$ to be reasonably large.

\bigskip

\begin{exercise}\label{ex:oneway}
 Show that the effective single-qubit operation corresponds to
\medskip
\[
\Qcircuit @C=0.7em @R=.7em @!R {
& \qw & \gate{H Z(\alpha)} &  \gate{H Z((-1)^k\beta)} & \gate{H Z((-1)^l\gamma)} & \qw & \gate{X^k} & \gate{Z^l} & \gate{X^m} & \qw
}
\]
Note that every measurement depends at most on the previous measurement outcome. How do the final three Pauli's affect the computation?
\end{exercise}

\begin{exercise}
 Find the action of the type-I fusion operator conditioned on detecting a vertically polarized photon. 
\end{exercise}

\section{Quantum computing with matter qubits and photons}\label{lec:dh}

At this point we have pretty much all the ingredients that we need for linear optical quantum computing. One important component is still missing, however. Because we rely on post-selection and feed-forward in this quantum computer architecture, we need the ability to store the qubits (the single photons) from the time they are first entangled with the cluster state, to the time they are detected. This means we need an optical {\em quantum memory}.

\subsection{Qubit memories}

Loosely speaking, let the {\em fault-tolerant threshold} be the maximum error beyond which no error correction can save the quantum computation. A quantum memory for linear optical quantum computing with single photons must then meet the following strict requirements:
\begin{enumerate}
	\item The photon must couple {\em into} the memory with high enough probability to surpass the fault-tolerant threshold.
	\item The photon must couple {\em out of} the memory with high enough probability to surpass the fault-tolerant threshold.
	\item The mode shape of the output photon must be identical to that of the input photon in order to facilitate high-fidelity interferometry. 
\end{enumerate}
Moreover, the memory errors are cumulative, so if all three errors above are just below the fault-tolerant threshold, the total error will surely be above the threshold.

A typical quantum memory used in experiments is a fibre-optical delay line. However, due to losses in the fibre this is not a scalable solution (in a full-scale quantum computer, the memory time is likely to be several clock cycles long). Therefore, for optical quantum computing with single photons to become a viable technology, we need some other system that can store the qubit value of the photon. Possibilities are atomic vapours, or systems with optical transitions strongly coupled to a cavity.

However, now we've just lost the advantage of the single photon as our qubit, namely its robustness against decoherence: The decoherence will now take place in the quantum memory. This means that we still have to create very robust matter qubits. So rather than trying to couple photons {\em into} the memory, we can engineer the capability of single-qubit operations into the memory and use the memories themselves as {\em matter qubits}. This removes requirement 1 above. In the next section, I will show that we can also remove requirement 2.

\subsection{The double-heralding protocol}

Since it seems that we need some matter system as a quantum memory for optical quantum computing, we will explore this avenue further and start out with the assumption that the memory is actually our qubit. The qubit can generate a photon depending on its state, and if we can apply fusion-style gates on two such photons, we may be able to create cluster states in matter qubits. Let's consider a matter system with two energy levels $|\down\rangle$ and $|\up\rangle$ that make up the qubit states. An excited level $|e\rangle$ in the system couples only to the $|\down\rangle$ level via an optical transition \cite{barrett}. 

\begin{figure}[t]
  \begin{center}
       \epsfig{file=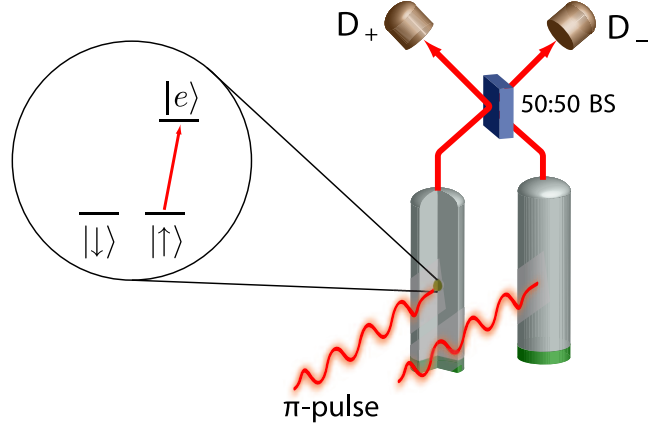, width=8.5cm}
       \caption{Schematic of the double-heralding procedure. The two qubits are in separate physical systems and interact solely through their emitted photons. Path erasure of the photons generates the entanglement between the qubits.}
  \end{center}
\end{figure}

We entangle two qubits by first preparing two of these systems in separate
cavities in the separable (unnormalised) state  $(|\up\rangle +
|\down\rangle)(|\up\rangle + |\down\rangle)$. Subsequently, we apply
an optical $\pi$-pulse to each system, and wait for a single photon to
be emitted. This yields the total state
\begin{equation}\nonumber |\up\up\rangle |0,0\rangle +
|\up\down\rangle |0,1\rangle + |\down\up\rangle |1,0\rangle +
|\down\down\rangle |1,1\rangle \; ,
\end{equation}  where $|0\rangle$ and $|1\rangle$ now denote the
vacuum and a single photon in the freely propagating optical mode
leaving the cavity, respectively. When these two modes interact on a
50:50 beam splitter, the total state becomes (note the HOM effect)
\begin{equation}\nonumber |\up\up\rangle |0,0\rangle +
\frac{1}{\sqrt{2}} \Bigl[(|\up\down\rangle + |\down\up\rangle)
|0,1\rangle \Bigr. + \Bigl. (|\up\down\rangle - |\down\up\rangle
|1,0\rangle + |\down\down\rangle  (|2,0\rangle + |0,2\rangle) \Bigr]
\; .
\end{equation}  Detecting both the outgoing modes of the beam
splitter, each with a {\em realistic} detector (i.e., a detector with
finite efficiency, and which cannot discriminate between optical
states with one or more photons), gives the following state of the
qubits (given just a single detector click in $D_{\pm}$):
\begin{equation}\label{intermediate}  \rho^{(\pm)} = f(\eta)
|\Psi^{(\pm)}\rangle\langle\Psi^{(\pm)}| + [1-f(\eta)]
|\down\down\rangle\langle\down\down|\; ,
\end{equation}  where $|\Psi^{(\pm)}\rangle = (|\up\down\rangle \pm
|\down\up\rangle)/\sqrt{2}$ and $f(\eta)\leq 1$ is a function of the
combined collection and detection efficiency, $\eta$. 

The state in Eq.~(\ref{intermediate}) is an incoherent mixture of a
maximally entangled state and the separable state
$|\down\down\rangle\langle\down\down|$. However, we can remove this
separable part by first applying a bit flip operation $|\up\rangle
\leftrightarrow |\down\rangle$ to both matter qubits.  We subsequently
apply a second $\pi$-pulse to each matter system. The separable part
cannot generate photons. Thus, conditional on observing another single
detector click, we obtain the final two-qubit pure state
\begin{equation}  |\Psi^{(\pm)}\rangle =
\frac{1}{\sqrt{2}}(|\up\down\rangle \pm |\down\up\rangle)\;
\end{equation}  The total success probability of this procedure is
$\eta^2/2$. Note that we have removed requirement 1 of the quantum memory since we do not couple photons {\em into} the matter system, and we alleviated requirement 2 of quantum memories by allowing for a reduced success probability of the entangling operation. The remaining challenge is to make indistinguishable the photons originating from different qubits. Recently, a group in Paris managed to control two atoms in optical tweezers sufficiently well so that the photons they emit are indistinguishable enough to show  two-photon quantum interference \cite{beugnon}.

\subsection{Creating cluster states with double-heralding}

The {\em double-heralding} entangling procedure described above is very similar to the type-II fusion gate, in that it effectively performs a projective parity measurement. Double heralding can therefore be used to create cluster states for universal quantum computing. Let's see how it works in detail.

We again use the formalism used in lecture \ref{lec:cluster}, where we write out the cluster state in terms of the conditional $Z$ operations. It is straightforward to show that the action of the double-heralding procedure is given by
\begin{equation}
	E_+ = |01\rangle\langle 01| + |10\rangle\langle 10| \quad\text{and}\quad
	E_- = |01\rangle\langle 01| - |10\rangle\langle 10|\; ,
	\label{eq:dhprojector}
\end{equation}
where we have identified $|\down\rangle$ with $|0\rangle$ and $|\up\rangle$ with $|1\rangle$, and the labels + and -- of the operator $E$ denote the detection signature.
Suppose we have two cluster states that we wish to join using the double-heralding procedure. Before they are connected, their state can again be written as 
\begin{equation}
	\left( |0\rangle_1 |\psi\rangle_a + |1\rangle_1  \prod_{j\in n(1)} {\mathcal{Z}}_j |\psi\rangle_a \right) \otimes \left( |0\rangle_2 |\phi\rangle_b + |1\rangle_2  \prod_{j\in n(2)} {\mathcal{Z}}_j |\phi\rangle_b \right) ,
	\label{eq:2clustersdh}
\end{equation}
with the qubit neighborhoods $n(j)$ defined as before. Applying the operator $E_+$ (i.e., a successful entangling operation) then yields the state
\begin{equation}
	|0\rangle_1 |1\rangle_2 |\psi\rangle_a \left( \prod_{j\in n(2)} {\mathcal{Z}}_j |\phi\rangle_b \right) + |1\rangle_1 |0\rangle_2 \left(\prod_{j\in n(1)} {\mathcal{Z}}_j |\psi\rangle_a\right) |\phi\rangle_b\; .
	\label{eq:aftereplus}
\end{equation}

\begin{figure}[t]
  \begin{center}
       \epsfig{file=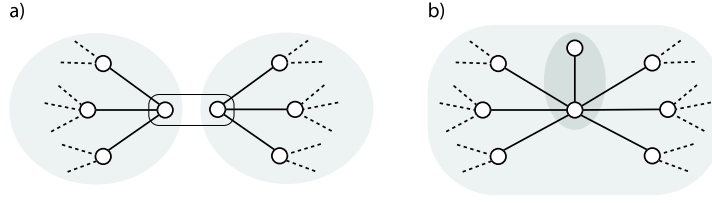, width=9.5cm}
     \caption{Joining to cluster states. a) The two separate clusters. b) The double-heralding operation creates a redundantly encoded qubit (the qubits in the dark shaded area).}\label{fig:join}
  \end{center}
\end{figure}

We need to show that this is again locally equivalent to a cluster state. To this end, apply a Hadamard operation $H_2$ to qubit 2 and a bit flip $X_1$ to qubit 1. Since both operators are part of the Clifford group, this will not destroy the cluster state:
\begin{eqnarray}
	|C\rangle &=& |00\rangle_{12} \left( \prod_{j\in n(1)} {\mathcal{Z}}_j |\psi\rangle_a \right) |\phi\rangle_b +  |01\rangle_{12} \left( \prod_{j\in n(1)} {\mathcal{Z}}_j |\psi\rangle_a \right) |\phi\rangle_b \cr && + |10\rangle_{12} |\psi\rangle_a \left( \prod_{j\in n(2)} {\mathcal{Z}}_j |\phi\rangle_b \right) - |11\rangle_{12} |\psi\rangle_a \left( \prod_{j\in n(2)} {\mathcal{Z}}_j |\phi\rangle_b \right)\; .
\end{eqnarray}
The question is: Is this another cluster state? In order to show that this is indeed the case, it is sufficient to show that we can transform it into a known form of a cluster state using local Clifford operations. So let's apply $\prod_j {\mathcal{Z}}_{j\in n(1)}$ to the qubits in set $a$. Since ${\mathcal{Z}}_j^2 = \unity$, we have 
\begin{eqnarray}
	|C'\rangle &=& |00\rangle_{12} |\psi\rangle_a |\phi\rangle_b + |10\rangle_{12} \left( \prod_{j\in n(1)} {\mathcal{Z}}_j |\psi\rangle_a \right) \left( \prod_{j\in n(1)} {\mathcal{Z}}_j |\phi\rangle_b \right) \cr && + |01\rangle_{12} |\psi\rangle_a |\phi\rangle_b - |11\rangle_{12} \left( \prod_{j\in n(2)} {\mathcal{Z}}_j |\psi\rangle_a \right) \left( \prod_{j\in n(2)} {\mathcal{Z}}_j |\phi\rangle_b \right)\; .
\end{eqnarray}
This can be written as
\begin{eqnarray}
	|C'\rangle &=& |0\rangle_1 |\Psi\rangle_c \left( |0\rangle_2 + |1\rangle_2 \right) + |1\rangle_1 \left( \prod_{l\in n(1)} {\mathcal{Z}}_l |\Psi\rangle_c \right) \left( |0\rangle_2 - |1\rangle_2 \right) \cr
	&=& |0\rangle_1 |\Psi\rangle_c |+\rangle_2 + |1\rangle_1 \left( \prod_{l\in n(1)} {\mathcal{Z}}_l  {\mathcal{Z}}_2 |\Psi\rangle_c |+\rangle_2 \right)\; .
	\label{eq:finaldh}
\end{eqnarray}
It is clear that this is again of the form of a cluster state, since qubit 2 has experienced a $Z$ operation depending on the state of qubit 1. We can in principle add qubit 2 to the set $c$ and expand the neighbourhood $n(1)$. However, leaving it in this form reveals something interesting about the cluster. Qubit 2 is not entangled with any qubit other than qubit 1. We call this a leaf or a cherry in the cluster. More accurately, qubit 1 and 2 form a {\em redundantly encoded qubit}, useful for error correction (see Fig.~\ref{fig:join}b).

\subsection{Complete quantum computer architecture}

Before I discuss the complete architecture of a quantum computer based on double heralding, let's explore some of the advantages and disadvantages of this approach.

The main advantage of the double-heralding protocol is that the resulting entanglement is completely independent of both the detector efficiency {\em and} the detector number-resolving capability. This is important, because it is extremely challenging to make photo-detectors with near perfect ($>98\%$) efficiency while keeping unwanted dark counts low. Because of this insensitivity to photon collection efficiency, it is also not necessary for the qubit to be in the strong coupling regime of the interaction between the optical transition and the electromagnetic field. Another advantage is that the protocol is inherently distributed: It does not matter whether the qubits are $10~\mu$m apart, or $10~$km. This is extremely useful for quantum communication. But more importantly, it allows us to really isolate each individual qubit and get a good control over decoherence. In addition, a slowly varying (random) phase in one of the input modes of the beam splitter will give at most an unobservable global phase shift. Finally, the protocol requires only a relatively simple level structure. There are potentially many systems that can be used for this scheme, from trapped ions and atoms, to NV centres in diamond and Pauli blockade quantum dots.

There are two main disadvantages to double-heralding based quantum computing. First, the success probability of the entangling operation is bounded by one half, and with photon loss the probability becomes $\eta^2/2$, where $\eta$ is the total photon collection efficiency. When the losses in the system are considerable, this makes the creation of cluster states a very costly affair (even though we maintain mathematical scalability at all times). Fortunately, there is a way to circumvent this problem and simultaneously keep the advantages of double heralding. It is called the {\em broker-client} model \cite{broker}. Instead of one qubit per site, we engineer two qubits with a high-fidelity, high efficiency (but non-scalable) two-qubit gate. An example of this is an NV center in diamond, where the two qubits are the electron spin and the nuclear spin. The nuclear spin is long-lived, and can be used to store a qubit from a cluster state. The electron spin can then be entangled with other electron spins via double-heralding, and when this succeeds, the entanglement is transferred to the nuclear spin using the two-qubit gate. This way, we can build up large cluster states without suffering exploding overhead costs.

The second disadvantage of this scheme is that the qubits must be almost identical. If they are not, the photons are likely to carry some information about their origins, and the entangling procedure gives us only non-maximal entanglement. In terms of the HOM experiment, the cancellation of detection coincidences at the output modes is no longer complete. When the photo-detectors have good time resolution, we can counter this problem to some degree \cite{campbell}: Knowledge of the arrival times of the photons in the detectors will ensure that the resulting entangled state remains pure, and a sophisticated adaptive strategy of which qubits to entangle next allows for some variation in the qubits.

\begin{figure}[t]
  \begin{center}
       \epsfig{file=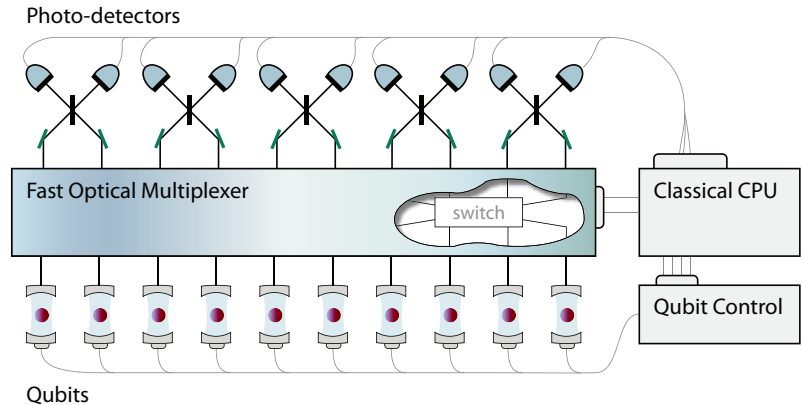, width=10.5cm}
     \caption{Schematic of a quantum computer using double heralding. Maximum parallelizability is obtained using a fast optical multiplexer. A classical CPU is needed for the tracking of the measurement outcomes, the driving of the qubit control, and the setting of the multiplexer.}\label{fig:qc}
  \end{center}
\end{figure}

How do we put all this together? Fig.~\ref{fig:qc} shows a quantum computer that operates on the double-heralding principle. It has five main components:

\begin{enumerate}
 \item The {\bf Qubits} are kept in individual environments in order to keep decoherence to a minimum. In the broker-client model, there may be multiple qubits per site. The qubits must be nearly identical in order to create high-fidelity cluster states.
 \item The {\bf Qubit control} component is designed to address the individual qubits, applying both the $\pi$-pulses, the bit flips, and the single-qubit rotations needed for the qubit measurements. This may involve multiple lasers and/or microwave fields.
 \item The {\bf Optical multiplexer} is a router that directs the optical output modes of the qubits into the beam splitters. This way, we can apply the double-heralding procedure to two arbitrary qubits in the quantum computer. The 50:50 beam splitters that are drawn outside the multiplexer in Fig.~\ref{fig:qc} can be incorporated as well, so that we can in principle do a complete readout of all qubits in one clock cycle.
 \item The {\bf Photo-detectors} must have reasonably high detection efficiency and very low dark count rate. Good time resolution is also an advantage. There is no need for single-photon resolution.
 \item The {\bf Classical CPU} keeps track of the measurement outcomes, controls the switching of the multiplexer, and tells the qubit controller what to do. In addition, the CPU is used to program the quantum computer, and it interprets the final qubit readout.
\end{enumerate}

\bigskip

\begin{exercise}
 Calculate the effect of an unknown phase shift in one of the input modes of the beam splitter in the double-heralding protocol. 
\end{exercise}

\begin{exercise}
 Calculate the effect of partial which-path erasure.
\end{exercise}

\begin{exercise}
 Verify the projective action of the double-heralding procedure.
\end{exercise}

\begin{exercise}
 When we fail to add a micro-cluster to a cluster, how do we retrieve the cluster state?
\end{exercise}

\section{Quantum computing with optical nonlinearities}\label{lec:kerr}

In the previous lecture we have seen how we can circumvent the need for quantum memories when we use material systems as qubits, together with a probabilistic entangling procedure. In this lecture, I show that we can obviate the need for quantum memories by choosing the right nonlinear interaction, which makes the entangling procedure (near) deterministic.

\subsection{Kerr nonlinearities}

\begin{figure}[t]
  \begin{center}
   \begin{psfrags}
    \psfrag{t}{$\theta$}
       \epsfig{file=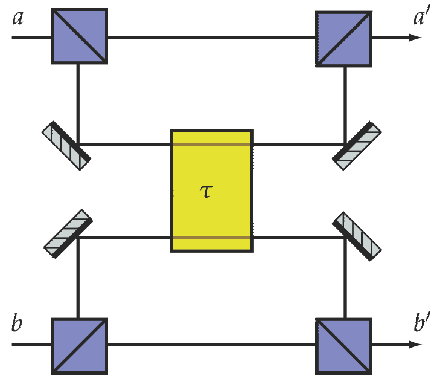, width=6cm}
   \end{psfrags}
  \end{center}
  \caption{A strong optical nonlinearity called a {\em cross-phase modulation} induces a phase shift on the vertically polarized part of mode $b$ that depends on the number of vertically polarized photons in mode $a$. The setup is symmetric and creates a {\sc cz} gate.}
  \label{fig:kerr}
\end{figure}

It has been known for a long time that we can make nonlinear optical gates using so-called Kerr nonlinearities. These consist of optically active materials that induce an effective photon-photon interaction. In particular, we consider the {\em cross-Kerr} nonlinearity shown in Fig.\ \ref{fig:kerr}. The central box has two input modes $a$ and $b$, and the interaction Hamiltonian is of the form
\begin{equation}
	H_K = \tau \hat{a}^{\dagger}\hat{a}\, \hat{b}^{\dagger}\hat{b}\; .
\end{equation}
This leads to the following Bogoliubov transformations on the annihilation operators:
\begin{equation}\label{eq:kerr}
	\hat{a} \rightarrow \hat{a}\, e^{i\tau \hat{b}^{\dagger}\hat{b}}, \qquad
	\hat{b} \rightarrow \hat{b}\, e^{i\tau \hat{a}^{\dagger}\hat{a}},
\end{equation}
in other words, the phase shift in mode $a$ depends on the intensity of the field in mode $b$.
It is straightforward to show that for $\tau = \pi$ and two polarized input photons in modes $a$ and $b$ respectively, Fig.\  \ref{fig:kerr} represents a {\sc cz} gate.

\begin{figure}[t]
  \begin{center}
  \begin{psfrags}
  \psfrag{a}{$\theta$}
  \psfrag{b}{$-\theta$}
  \psfrag{c}{$2\phi(x)$}
  \psfrag{d}{$|\alpha\rangle$}
  \psfrag{e}{$|\psi_{ab}\rangle$}
  \psfrag{f}{$|q_2\rangle$}
       \epsfig{file=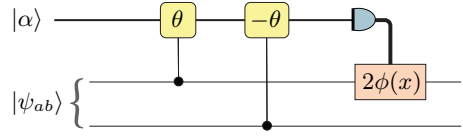, width=6cm}
	\end{psfrags}
  \end{center}
  \caption{A coherent state couples to two photonic qubits via the weak cross-Kerr nonlinearity indicated by $\theta$. The measurement outcome determines a phase shift $2\phi(x)$ on one qubit.}
  \label{fig:weak-setup}
\end{figure}

Unfortunately, there are no real materials that have the properties that $\tau = \pi$ and are otherwise free of noise. The question thus arises: What can we do if $\tau = \theta \ll \pi$? The answer is that we can again construct a parity gate \cite{barrett2}. To this end, we make use of a reasonably bright coherent state that will carry the quantum correlations from one photon to the other. We consider the setup in Fig.\  \ref{fig:weak-setup}. Let the two-qubit input state be $|\psi_{ab}\rangle = c_{00} |00\rangle + c_{01} |01\rangle + c_{10} |10\rangle + c_{11} |11\rangle$, and the coherent state is denoted by $|\alpha\rangle$. Assume that the interactions take place between $|1\rangle$ and $|\alpha\rangle$. The interaction is again a cross-Kerr nonlinearity, which produces a phase shift $\theta$ in the coherent state depending on the photon state in the signal mode. After the first interaction we obtain the three-mode optical state
\begin{equation}
	|\psi_1\rangle = c_{00} |00\rangle|\alpha\rangle + c_{01} |01\rangle|\alpha\rangle + c_{10} |10\rangle|\alpha\, e^{i\theta}\rangle + c_{11} |11\rangle|\alpha\, e^{i\theta}\rangle\; ,
\end{equation}
and after the second interaction we have 
\begin{equation}\label{psi2}
	|\psi_2\rangle = c_{00} |00\rangle|\alpha\rangle + c_{01} |01\rangle|\alpha\, e^{-i\theta}\rangle + c_{10} |10\rangle|\alpha\, e^{i\theta}\rangle + c_{11} |11\rangle|\alpha\rangle\; .
\end{equation}
We can separate this state into an even ($\{|00\rangle,|11\rangle\}$) and an odd ($\{|01\rangle,|10\rangle\}$) parity contribution. 

\begin{figure}[t]
  \begin{center}
   \begin{psfrags}
       \psfrag{s2}{$|\alpha\, e^{\pm i\theta}\rangle$}
       \psfrag{s1}{$|\alpha\rangle$}
       \psfrag{a}{a) Phase space}
       \psfrag{t}{$\theta$}
       \psfrag{x}{$x$}
       \psfrag{b}{b) Probability $p(x)$}
       \epsfig{file=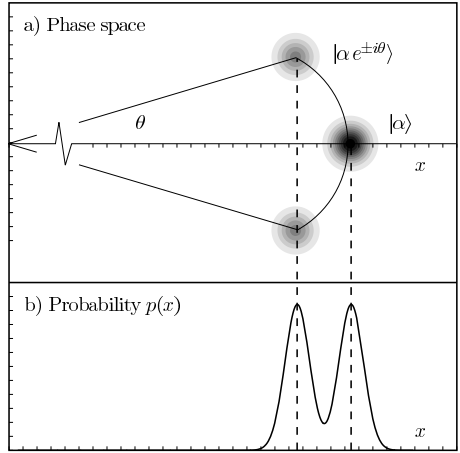, width=6cm}
   \end{psfrags}
  \end{center}
  \caption{a) Phase space representation of the weak nonlinear parity gate. b) The corresponding probability of the outcomes of an $x$-quadrature measurement. The overlap between the two Gaussian peaks must be made sufficiently large for the gate to work near deterministically.}
  \label{fig:weak}
\end{figure}

The next step is to measure the $x = (\hat{a} + \hat{a}^{\dagger})/\sqrt{2}$ quadrature of the coherent state. We see from Eq.~(\ref{psi2}) that such a measurement leaves the even parity subspace invariant, but not the odd subspace. To demonstrate a parity gate we calculate the projection of $|\psi_2\rangle$ onto the eigenstate of the measurement outcome $|x\rangle$:
\begin{equation}
	\langle x|\psi_2\rangle = \left( c_{00} |00\rangle + c_{11} |11\rangle \right) \langle x|\alpha\rangle + c_{01} |01\rangle \langle x|\alpha\, e^{-i\theta}\rangle + c_{10} |10\rangle \langle x|\alpha\, e^{i\theta}\rangle \; .
\end{equation}
Using Eq.~(A4.12) on page 235 of Ref.~\cite{gz}:
\begin{equation}
	\langle x|\alpha\rangle = \frac{1}{\sqrt[4]{\pi}} \exp\left[ -\frac{1}{2} \left( x - \sqrt{2}\alpha \right)^2 + \frac{1}{2} \alpha\left( \alpha - \alpha^* \right) \right] ,
\end{equation}
and assuming that $\alpha$ is real\footnote{In addition, we describe the coherent state in the co-rotating frame of reference, which allows us to suppress the free time evolution of the coherent state. In particular, this means that $\alpha$ is real for all times, and the nonlinear phase shift $\theta$ is included explicitly.}, we find that 
\begin{equation}
	\langle x|\alpha\rangle = \frac{1}{\sqrt[4]{\pi}} \exp\left[ -\frac{1}{2} \left( x - \sqrt{2}\alpha \right)^2 \right]
\end{equation}
and 
\begin{equation}
	\langle x|\alpha e^{i\theta}\rangle = \frac{1}{\sqrt[4]{\pi}} \exp\left[ -\frac{1}{2} \left( x - \sqrt{2}\alpha\, \cos\theta \right)^2 + i \alpha\, \sin\theta\, (\sqrt{2} x - \alpha\,\cos\theta) \right].
\end{equation}
The state after the measurement is therefore 
\begin{equation}\label{eq:wk}
	|\psi_{ab}'\rangle = \langle x|\alpha\rangle \left( c_{00} |00\rangle + c_{11} |11\rangle \right) + |\langle x|\alpha\, e^{i\theta}\rangle| \left( c_{01}\, e^{-i\phi} |01\rangle + c_{10}\, e^{i\phi} |10\rangle \right)\; ,
\end{equation}
with
\begin{equation}\label{eq:pswk}
	\phi(x) \equiv \alpha\, \sin\theta\, (\sqrt{2} x - \alpha\,\cos\theta)\; .
\end{equation}
A phase space representation of the above procedure is given in Fig.\  \ref{fig:weak}. The relative phase $2\phi(x)$ can be corrected using regular phase shifts.

These two distributions peak at different values $x_e$ and $x_o$ for the even and odd subspace respectively:
\begin{equation}
	x_e = \sqrt{2}\alpha \qquad\text{and}\qquad x_o = \sqrt{2}\alpha\,\cos\theta\;.
\end{equation}
The width of (the real part of) these distributions is of the order one. We can distinguish the two peaks (and correspondingly obtain a high fidelity) when $x_e - x_o$ is larger than twice the width of the distribution $(\theta \ll 1)$:
\begin{eqnarray}
	&& x_e - x_o > 2 \quad \Leftrightarrow \quad \sqrt{2}\alpha\, (1-\cos\theta) > 2
	\quad \Leftrightarrow \quad \alpha\, \frac{\theta^2}{2\sqrt{2}} > 1 \; .
\end{eqnarray}
Weak nonlinearities on the order of $10^{-5}$ can be achieved using electromagnetically induced transparencies. There are several tricks that can be used to increase the performance of this gate \cite{spiller}.

An $x$-quadrature measurement that can project the two-photon state onto either one of the parity subspaces is again a parity gate, and we have seen earlier how these projections are useful for quantum computing. Here, the parity projection is practically deterministic if the peak separation is big enough, which means that cluster state growth can be very efficient.

\subsection{Zeno gates}

\begin{figure}[t]
  \begin{center}
       \epsfig{file=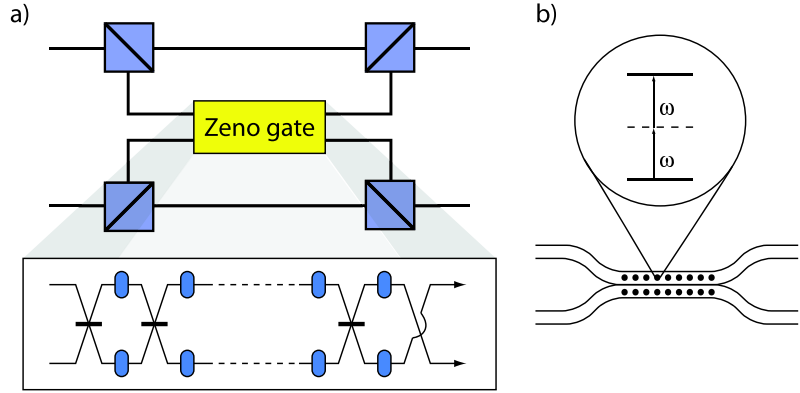, width=10.5cm}
  \end{center}
  \caption{(a) A {\sc cz} gate using the Zeno effect. The $n$ beam splitters have transmittivity $1/n$, and are separated by dissipative two-photon absorbers. In the limit of $n \rightarrow \infty$, and perfect two-photon absorption, the Zeno gate implements a perfect {\sc cz} gate. (b) The beam splitters are implemented using a spliced fibre with two cores, each filled with atoms that absorb two-photon excitations.}
  \label{fig:zeno}
\end{figure}

Another optical nonlinearity that may be used to construct a near-deter\-minis\-tic two-photon gate is two-photon absorption. This is the basis of the so-called Zeno gate by Franson, Jacobs, and Pittman \cite{franson}, and is shown in Fig.~\ref{fig:zeno}. The gate works similarly to the strong Kerr gate shown in Fig.~\ref{fig:kerr}, but the detailed physics of the central (yellow) box differs. 

Before I describe the Zeno gate, let's look at a possible experimental implementation of a beam splitter: Typically, we think of a beam splitter as a semi-reflective mirror, but there are also other ways. Many people who study photons in optical fibres make the beam splitters with fibres as well, so how is this done? In general, {\em any} unitary two-mode transformation can be described by the matrix given in Eq.~(\ref{2bs}). If we take a length of fibre and splice it at both ends, we end up with two fibres that join for a certain length, and then separate again. Because the action of such a physical object behaves according to Eq.~(\ref{2bs}), we can model this as a beam splitter, where the transmission coefficient is now related to the length of the joined piece of fibre. The evolution is therefore something like this:
\begin{eqnarray}
	|00\rangle &\rightarrow& |00\rangle \cr
	|01\rangle &\rightarrow& \cos\theta |01\rangle + i\sin\theta |10\rangle \cr
	|10\rangle &\rightarrow& i\sin\theta |01\rangle + \cos\theta |10\rangle \cr
	|11\rangle &\rightarrow& \cos 2\theta |11\rangle + \frac{i}{\sqrt{2}} \left( |20\rangle + |02\rangle \right),
	\label{eq:zeno}
\end{eqnarray}
where 0, 1, and 2 denote the photon occupation number of the mode, and we have chosen a convenient phase convention\footnote{We can always include phase shifts in the fibres to make the interaction of this form.}.

The Zeno gate works like this spliced fibre, with a small but essential modification: Consider two fibres with a hollow core that come together at some point, remain parallel for a certain distance, and then separate. Again, this is properly described by Eqs.~(\ref{2bs}) and (\ref{eq:zeno}). Photons entering one fibre couple to the other fibre via the evanescent electromagnetic field, and they can tunnel from one core to the other. If the length is chosen correctly, we can make the photons come out in the other fibre. When two photons enter the device, one in each input fibre, we can also demonstrate the Hong-Ou-Mandel effect\footnote{For the HOM effect to take place, the length of joined fibre must be half the length of the fibre in the case where single photons enter one core and exit the other.}. 

\bigskip

To make the Zeno gate, we fill both cores with a linear array of atoms that have a strong two-photon absorption and negligible single-photon absorption. When only a single photon enters the device, the atoms have no effect on the dynamics, and the photon exits in some superposition of the output modes. On the other hand, when two photons enter the device, one in each input mode, they are prevented from building up the two-photon amplitude due to the absorption: After a very short distance, the two-photon input state evolves according to Eq.~(\ref{eq:zeno}) with $\theta \ll 1$. When the photons encounter the first atom, the term in the superposition with two photons in one mode will transform into the vacuum because the atom absorbs the photons and dissipates the energy into the environment. This is effectively a measurement where we throw away the measurement outcome.

Since the length of free evolution in the fibre cores is so short (the atoms are placed closely together), the amplitude $i\sin\theta$ is very small, and the probability of two-photon absorption is also tiny. Therefore, after the photons encounter the atoms, they are projected onto the $|11\rangle$ state with very high probability. En route to the next atom, they will evolve again, and this procedure repeats until the fibre cores separate. The atoms act as an almost continuous measurement, preventing the state from building up an appreciable absorption amplitude. This is commonly known as the {\em Zeno} effect. The photons will exit the interaction region in different modes, due to the suppression of the HOM effect. 

Why does this work as a two-photon gate? To answer this, we look at the accumulated phases of the four possible input states $|00\rangle$, $|01\rangle$, $|10\rangle$, and $|11\rangle$. Clearly, the state $|00\rangle$ remains unchanged, because there are no photons at all. The length of the interaction is chosen such that a single-photon input ($|01\rangle$ or $|10\rangle$) is transmitted perfectly into the other core. The accumulated phase for a transmitted photon is $e^{i\pi/2}$, according to Eq.~(\ref{eq:zeno}). Finally, when two photons enter the device (i.e., the state $|11\rangle$), the beam splitter action is suppressed, and the photons are effectively reflected. The phase associated with perfect reflection is 1. By choosing suitable phase shifts in the output modes, this interaction can be turned into a {\sc cz} gate:
\begin{eqnarray}
	U_{\rm Zeno} = 
	\begin{pmatrix}
	 1 & 0 & 0  & 0 \cr
	 0 & i & 0  & 0 \cr
	 0 & 0 & i  & 0 \cr
	 0 & 0 & 0  & 1 \cr
	\end{pmatrix}
	\qquad\rightarrow_{\rm phase~shift}\qquad
  U_{CZ} = 
	\begin{pmatrix}
	 1 & 0 & 0  & 0 \cr
	 0 & 1 & 0  & 0 \cr
	 0 & 0 & 1  & 0 \cr
	 0 & 0 & 0  & -1 \cr
	\end{pmatrix}.
	\label{eq:zenocz}
\end{eqnarray}
This is the physical intuition behind the Zeno gate. Let's derive this result slightly more formally. 

We consider the ideal case where the two-photon absorption is perfect, and there is no single-photon absorption (or loss). Since the two-photon absorption is followed by spontaneous emission into the environment, the evolution is incoherent, and we can no longer use a pure state description of the situation. We therefore construct the Positive Operator Valued Measures ({\sc povm}s) for the different measurement outcomes \cite{krauss}. In general, an arbitrary input state $\rho$ will evolve according to 
\begin{equation}
	\rho ~\rightarrow~ \tilde{\rho} = {\mathcal{L}}(\rho) \equiv \sum_{k=1,2} A_k \rho A_k^{\dagger},
	\label{eq:povm}
\end{equation}
where the $A_k$ are the Kraus operators (or {\em effects}) that define the effect of the measurement on the state. Each measurement outcome is represented by a specific $A_k$. Since we discard the measurement outcomes in our Zeno gate, we need to sum over all $k$. Note that here we are talking about the state $\rho$ of one optical mode (or one fibre core). The Kraus operators satisfy the relation $\sum_k A_k^{\dagger} A_k = \unity$, which ensures that $\tilde{\rho}$ is a proper density operator.

In this case we have two Kraus operators: one when there is no absorption, and one for two-photon absorption. When there is no absorption, nothing happens, and the corresponding Kraus operator is the {\em identity} operator on the relevant subspace (spanned by $|0\rangle\langle 0|$ and $|1\rangle \langle 1|$). On the other hand, 
two-photon absorption can be formalized as changing the state $|2\rangle$ into $|0\rangle$. We therefore have
\begin{equation}
	A_1 = |0\rangle\langle 0| + |1\rangle\langle 1| \quad\text{and}\quad
	A_2 = |0\rangle\langle 2|.
	\label{eq:krauss}
\end{equation}
In order to evaluate the effect of the Zeno gate we need to apply the super-operator ${\mathcal{L}}(\rho)$ to both modes every time an atom is encountered. Clearly, ${\mathcal{L}}(\rho)$ is acting as the identity if there is at most one photon in the system, so $|01\rangle\rightarrow i |10\rangle$ and $|10\rangle\rightarrow i |01\rangle$. But what about the $|11\rangle$ term?

The density operator for the state $|11\rangle$ is given by $|11\rangle\langle 11|$, and the beam splitter evolution in Eq.~(\ref{eq:zeno}) will give
\begin{eqnarray}
	\rho &=& \cos^2 2\theta |11\rangle\langle 11| + \frac{1}{2} \sin^2 2\theta \left(  |20\rangle + |02\rangle \right) \left( \langle 20| + \langle 02| \right) \cr
	&& + \frac{i}{\sqrt{2}} \cos 2\theta\sin 2\theta \left[ \left( |20\rangle + |02\rangle \right) \langle 11| - |11\rangle \left( \langle 20| + \langle 02| \right) \right] .
	\label{eq:mixed evolution}
\end{eqnarray}
When we apply the super-operator ${\mathcal{L}}(\rho)$, we find that 
\begin{equation}
	\tilde{\rho} = \sum_k A_k \rho A_k^{\dagger} = \cos^2 2\theta |11\rangle \langle 11| + \sin^2 2\theta |00\rangle \langle 00|.
	\label{eq:super}
\end{equation}
The term $|00\rangle\langle 00|$ is invariant under both the beam splitter evolution and the two-photon absorption, and does not change during the remainder of the gate. The $|11\rangle \langle 11|$ term will again undergo the evolution in Eq.~(\ref{eq:super}). After the full length of the joined fibre (involving $n$ atoms), the evolution is
\begin{equation}
	|11\rangle \langle 11| ~\rightarrow~ \cos^{2n} 2\theta |11\rangle \langle 11| + (1-\cos^{2n} 2\theta) |00\rangle \langle 00| .
	\label{eq:}
\end{equation}

The case of the input state $|01\rangle$ is symmetric to the input state $|10\rangle$, so we need to discuss only one of them here. Since the Kraus operator in this subspace is the identity operator, the evolution is a series of $n$ rotations over angle $\theta$, which can be written as
\begin{eqnarray}
	|01\rangle &\rightarrow& \cos n\theta |01\rangle + i\sin n\theta |10\rangle , \cr
	|10\rangle &\rightarrow& i\sin n\theta |01\rangle + \cos n\theta |10\rangle .
	\label{eq:nrotation}
\end{eqnarray}
These are all the ingredients we need to analyze the ideal Zeno gate. 

Remember that for the Zeno gate to work the single-photon input states must be swapped (perfect transmission), while total reflection must occur when there are two photons entering the device, one in each input mode. Therefore, we must choose $n\theta = \pi/2$, and this generates a phase shift $i$ on the photon.

Using this choice of $\theta$, the probability amplitude of the $|11\rangle$ term becomes $\cos^n(\pi/n)$. In order to have a proper Zeno effect, $n$ must be very large. We can expand the cosine function to first order and take the limit of $n$ to infinity:
\begin{equation}
	\lim_{n\rightarrow\infty} \left( 1-\frac{\pi^2}{2n^2} \right)^n = 1.
	\label{eq:limit}
\end{equation}
Indeed, the two-photon absorption [$1-\cos^{2n}(\pi/n)$] is completely suppressed. Furthermore, the phase of the $|11\rangle$ term is unaffected. 

We now have the situation where single photons are transmitted (and accumulate a phase $i$), while two input photons are both reflected (and don't experience a phase shift). A simple swap of the output modes will then result in $|01\rangle \leftrightarrow |10\rangle$, and the transformation becomes of the form of $U_{\rm Zeno}$ in Eq.~(\ref{eq:zenocz}).

So far, we analyzed the Zeno gate in the ideal case of perfect two-photon absorption and no single-photon absorption using {\sc povm}s. When the situation is not ideal (e.g., in the case of survival of the $|20\rangle$ and $|02\rangle$ terms and photon loss), the Kraus operators need to be modified and the calculation will become much harder. Alternatively, the problem can be formulated in Lindbladt form or in terms of a master equation, which can then be solved using standard techniques \cite{gz}.

\bigskip

\begin{exercise}
 Prove Eq.~(\ref{eq:kerr}).
\end{exercise}

\begin{exercise}
 Verify Eqs.~(\ref{eq:wk}) and (\ref{eq:pswk}).
\end{exercise}

\begin{exercise}
 Calculate the fidelity of this gate for an equal-superposition input state $c_{00} = c_{01} = c_{10} = c_{11} = \frac{1}{2}$.
\end{exercise}

\begin{exercise}
 Verify Eq.~(\ref{eq:zeno}) and Eq.~(\ref{eq:zenocz}).
\end{exercise}

\begin{exercise}
 Check that the Kraus operators in Eq.~(\ref{eq:krauss}) obey the normalization condition, and verify Eq.~(\ref{eq:super}).
\end{exercise}

\section*{Final remarks}

Single photons are very resilient to decoherence, and they travel at very high speed. This makes them the ideal carriers for quantum information. It is therefore likely that optical systems will play an important role in future quantum information technology. However, the lack of a direct interaction between photons means that some trickery must be used if you want them to carry out quantum computations. There have been several proposals for optical quantum computers, the best known of which is the Knill-Laflamme-Milburn scheme using only photons, linear optics, and photo-detectors. This scheme, and its improvements, needs quantum memories, because they rely critically on the feed-forward of measurement outcomes to modify subsequent interferometry. As a consequence, the advantage of photons as slow-decohering qubits is lost: Decoherence now takes place in the memory.
 
In these lectures, I have argued that you can either choose to live with it and make the quantum memories your qubits (which of course means that you have to engineer high-quality qubits), or you can turn to nonlinear interactions to create deterministic gates (an equally daunting task). At this point it is not clear what the greater challenge is. Finally, all the results presented here rely to a greater or lesser extent on the ability to create identical single-photon wave packets.

\section*{Acknowledgments}

I thank Simon Benjamin and Dan Browne for valuable discussions, and Erik Gauger for carefully reading the manuscript. I also thank Francesco Petruccione and the University of Kwazulu-Natal in South Africa for inviting me to give these lectures. This work was done as part of the QIP IRC www.qipirc.org (GR/S82176/01).

\end{document}